\documentclass[12pt]{iopart}

\usepackage{amsfonts}
\usepackage{amssymb}
\usepackage{epsfig}
\usepackage{graphicx}

\usepackage{epsf}

\usepackage{epstopdf}

\newcommand{\kb}{k_{\mathrm{B}}}
\newcommand{\nuc}{\nu_{\mathrm{c}}}
\newcommand{\nuw}{\nu_{\mathrm{w}}}
\newcommand{\nub}{\nu_{\mathrm{b}}}
\newcommand{\rhom}{\rho_{\mathrm{m}}}
\newcommand{\rhof}{\rho_{\mathrm{f}}}
\newcommand{\rhoc}{\rho_{\mathrm{c}}}
\newcommand{\ud}{\/\mathrm{d}\/}
\newcommand{\ple}{P^{(\mathrm{leq})}}
\newcommand{\pmic}{P^{(\mathrm{eq})}}
\newcommand{\hks}{h_{\mathrm{KS}}}
\newcommand{\Pro}{\mathrm{Prob}}
\newcommand{\pple}{p^{(\mathrm{leq})}}

\begin{document}
\paper[Fourier's law in many particle dispersing billiards]
{Heat conduction and Fourier's law in a class of
   many particle dispersing billiards}
\author{Pierre Gaspard\textdagger, Thomas Gilbert\textdaggerdbl}
\address{Center for Nonlinear Phenomena and Complex Systems,
   Universit\'e Libre  de Bruxelles, C.~P.~231, Campus Plaine, B-1050
   Brussels, Belgium}
\date{\today}
\begin{abstract}
We consider the motion of many confined billiard balls in interaction and
discuss their transport and chaotic properties. In spite of the absence
of mass transport, due to confinement, energy transport can take place
through binary collisions between neighbouring particles. We explore the
conditions under which relaxation to local equilibrium occurs on time
scales much shorter than that of binary collisions, which characterize the
transport of energy, and subsequent relaxation to local thermal equilibrium.
Starting from the pseudo-Liouville equation for the time evolution of
phase-space distributions, we derive a master equation
which governs the energy exchange between the system constituents. We
thus obtain analytical results relating the transport coefficient of
thermal conductivity to the frequency of collision events and compute these
quantities. We also provide estimates of the Lyapunov exponents and
Kolmogorov-Sinai entropy under the assumption of scale separation. The
validity of our results is confirmed by extensive numerical studies.
\end{abstract}
\submitto{\NJP}
\ead{\textdagger gaspard@ulb.ac.be, \textdaggerdbl thomas.gilbert@ulb.ac.be}
\maketitle

\section{Introduction}

Understanding the dynamical origin of the mechanisms which underly the
phenomenology of heat conduction has remained one of the major open
problems of statistical mechanics ever since Fourier's seminal work
\cite{F1822}. Fourier himself actually warned his reader that the effects
of heat conduction ``make up a special range of phenomena which cannot be
explained by the principles of motion and equilibrium,'' thus seemingly
rejecting the possibility of a fundamental level of description. Of course,
with the subsequent developments of the molecular kinetic theory of heat,
starting with the works of the founding-fathers of statistical mechanics,
Boltzmann, Gibbs and Maxwell, and later with the definite triumph of
atomism, thanks to Perrin's 1908 measurement of the Avogadro number in
relation to Einstein's work on Brownian motion, Fourier's
earlier perception was soon discarded as it became clear that heat transport
was indeed the effect of mechanical causes.

Yet, after well over a century of hard labor, the community is still
actively hunting for a first principles based derivation of Fourier's law,
some authors going so far as to promise ``a bottle of very good wine to
anyone who provides [a satisfactory answer to this
challenge]'' \cite{BLRB00}. Thus the challenge is, starting from the
Hamiltonian time evolution of a system of interacting particles which
models a fluid or a crystal, to derive the conditions under which the heat
flux and temperature gradients are linearly related by the coefficient of
heat conductivity. This is the embodiment of Fourier's law.

As outlined in \cite{BLRB00}, it is necessary, in order to achieve this,
that the model statistical properties be fully determined in terms of the
local temperature, a notion which involves that of local thermalization. To
visualize this, we imagine that the system is
divided up into a large number of small volume elements, each large enough
to contain a number of particles whose statistics is accurately described
by the equilibrium statistics at the local temperature associated to the
volume element under consideration. To establish this property, one must
ensure that time scales separate, which implies that the volume elements
settle to local thermal equilibrium on time scales much shorter than the
ones which characterize the transport of heat at the macroscopic
scale. Local thermal equilibrium therefore relies on very strong ergodic
properties of the model.

The natural framework to apply this programme is that of chaotic
billiards. Indeed non-interacting particle billiards, where individual
tracers move and make specular collisions among a periodic array of fixed
convex scatterers (the periodic Lorentz gases) are the only known
Hamiltonian systems for which mass transport \cite{BS80} and shear and bulk
viscosities \cite{BS96} have been established in a rigorous way.
Here, rather than local thermal equilibrium, it is a relaxation to local
equilibrium, occurring on the constant energy sheet of the individual
tracer particles, which allows to model the mass transport by a random walk
and yields an analytic estimate of the diffusion coefficient
\cite{MZ83}. Arguably Lorentz gases, whether periodic or disordered, in or
out of equilibrium, have played, over more than a century, a privileged and
most important role in the development of transport and kinetic theories
\cite{GND03}. However, in the absence of interaction among the tracer
particles, there is no mechanism for energy exchange and therefore no
process of thermalization before heat is conducted. If, on the other hand,
one adds some interaction between the tracers, say, as suggested in
\cite{BLRB00},
by assigning them a size, the resulting system will typically have
transport equations for the diffusion of both mass and energy. Though these
systems may be conceptually simple, they are usually mathematically too
difficult to handle with the appropriate level or rigor.

An example of billiard with interacting tracer particles is the
modified Lorentz gas with rotating discs proposed by Mej\'{\i}a--Monasterio
{\em et al.} \cite{MLL01}. This system exhibits normal transport with
non-trivial coupled equations for heat and mass transport. Though this
model has attracted much attention among mathematicians
\cite{EY04,EY06,EMMZ06,LY07,RY07,EJ07},
the rigorous proof of Fourier's law for such a system of arbitrary size and
number of tracers relies on assumptions which, though they are plausible, are
themselves not resolved. 
Moreover the equations which
govern the collisions and the discs rotations do not, as far as we know,
derive from a Hamiltonian description.

Yet a remedy to such limitations with respect to the number of
particles in interaction that a rigorous treatment will handle may have
been just around the corner \cite{K07}, and, this, within a Hamiltonian
framework. Indeed in \cite{BLPS92}, Bunimovich {\em et al.} introduced a
class of dispersing billiard tables with particles that are in geometric
confinement --{\em   i.~e.} trapped within cells-- but that can nevertheless
interact among particles belonging to neighbouring cells. The authors proved
ergodicity and strong chaotic properties of such systems with arbitrary
number of particles.

More recently, the idea that energy transport can be modeled as a slow
diffusion process resulting from the coupling of fast energy-conserving
dynamics has led to proofs of central limit theorems in the context of
models of random walks and coupled maps which describe the diffusion of
energy in a strongly chaotic, fast changing environment
\cite{DL07,BK08}. Although the extension of these results to symplectic
coupled maps, let alone Hamiltonian flows, is not yet on the horizon, it
is our belief that such systems as the dispersing billiard tables
introduced in \cite{BLPS92} will, if any, lend
themselves to a fully rigorous treatment of heat transport within a
Hamiltonian framework. As we announced in \cite{GG08a}, the reason why we
should be so hopeful is that the 
particles confinement has two important consequences~: first, relaxation to
local thermal equilibrium is preceded by a relaxation of individual
particles to local equilibrium\footnote{Let us underline the distinction
  we make here and in the sequel between relaxation to local equilibrium,
  which precedes energy exchanges, and relaxation to local thermal
  equilibrium, which involves energy exchanges among particles belonging to
  neighbouring cells.}, 
which occurs at constant energy within each 
cell, and has strong ergodic properties that guarantee the rapid decay of
statistical correlations; and, second, heat transport, unlike in the
rotating discs model, can be controlled by the mere geometry of the
billiard, which also controls the absence of mass transport. As of the
first property, the relaxation to a local equilibrium before energy
exchanges take place is characterized by a fast time scale, much faster
than that of relaxation to local thermal equilibrium among neighbouring
cells, which is itself much faster than the hydrodynamic relaxation
scale. There is therefore a hierarchy of three separate time scales in this
system, the first accounting for relaxation at the microscopic scale of
individual cells, the second one at the mesoscopic scale of neighbouring
cells, and the third one at the macroscopic scale of the whole system.

In this paper, we achieve two
important milestones towards a complete first principles derivation of the
transport properties of such models. Having defined the model, we establish
the conditions for separation of time scales and relaxation to local
equilibrium, identifying a critical geometry where binary collisions become
impossible. Assuming relaxation to local equilibrium holds, we go on to
considering the time evolution of phase-space densities and derive, from
it, a master equation which governs the exchange of energy in the system
\cite{MMN84}, thus going from a microscopic scale description of the
Liouville equation to the mesoscopic scale at which energy transport takes
place. We regard this as the first milestone, namely identifying the
conditions under which one can rigorously reduce the level of descrition
from the deterministic dynamics at the microscopic level to a stochastic
process described by a master equation at the mesoscopic level of energy
exchanges. 

This master equation is then used to compute the frequency of binary 
collisions and to derive Fourier's law and the macroscopic heat equation,
which results from the application of a small temperature gradient between
neighbouring cells. This is our second milestone~: an analytic formula 
for heat conductivity, exact for the stochastic system, and thus valid for
the determinisitc system at the critical geometry limit. These results are
then checked against numerical computations of these quantities, with
outstanding agreement, and shown to extend beyond the critical geometry,
with very good accuracy, to a wide range of parameter values. 

We further characterize the chaotic
properties of the model and offer arguments to account for the spectrum of
Lyapunov exponents of the system, as well as the Kolmogorov-Sinai entropy, 
expressions which are exact at the critical geometry. Again, these results
are very nicely confirmed by our numerical computations.

The paper is organized as follows. The models, which we coin lattice
billiards, are introduced in section \ref{sec.lb}. Their main geometric
properties are established, distinguishing transitions between insulating
and conducting regimes
under the tuning of a single parameter. The same parameter controls the
time scales separation responsible for local equilibrium. Section
\ref{sec.kt} provides the derivation of the master
equation which, under the assumption of local equilibrium, governs
energy transport. The main observables are computed and their scaling
properties discussed. In section \ref{sec.nr} we review the properties of
the model and assess the validity of the results of section \ref{sec.kt}
under the scope of our numerical computations.  The chaotic properties of
the models are discussed in section \ref{sec.ls}. We use simple theoretical
arguments to predict some of these properties and compare them to numerical
computations. Finally, conclusions are drawn in section \ref{sec.con}.

\section{Lattice billiards \label{sec.lb}}

To introduce our model, we start by considering the uniform motion of a
point particle about a dispersing billiard table, $\mathcal{B}_\rho$,
defined by the domain exterior to four overlapping discs of radii
$\rho$, centered at the four corners of a  square of sides $l$.
The radius is thus restricted to the interval
$l/2 \le \rho < l/\sqrt{2}$, where the lower bound is the overlap (or
bounded horizon) condition, and the upper bound is reached when
$\mathcal{B}_\rho$ is empty.

\begin{figure}[htb]
   \centerline{
     \includegraphics[width=.3\textwidth]{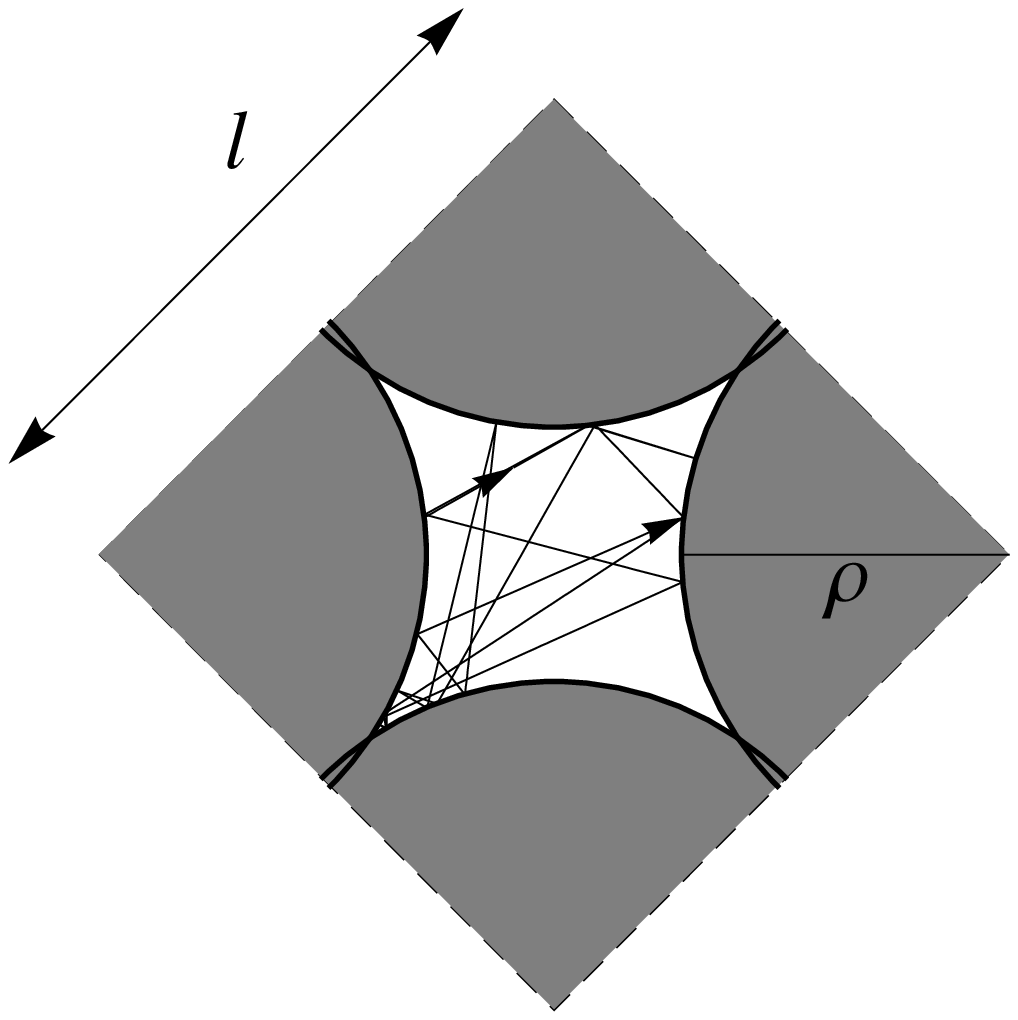}
     \includegraphics[width=.3\textwidth]{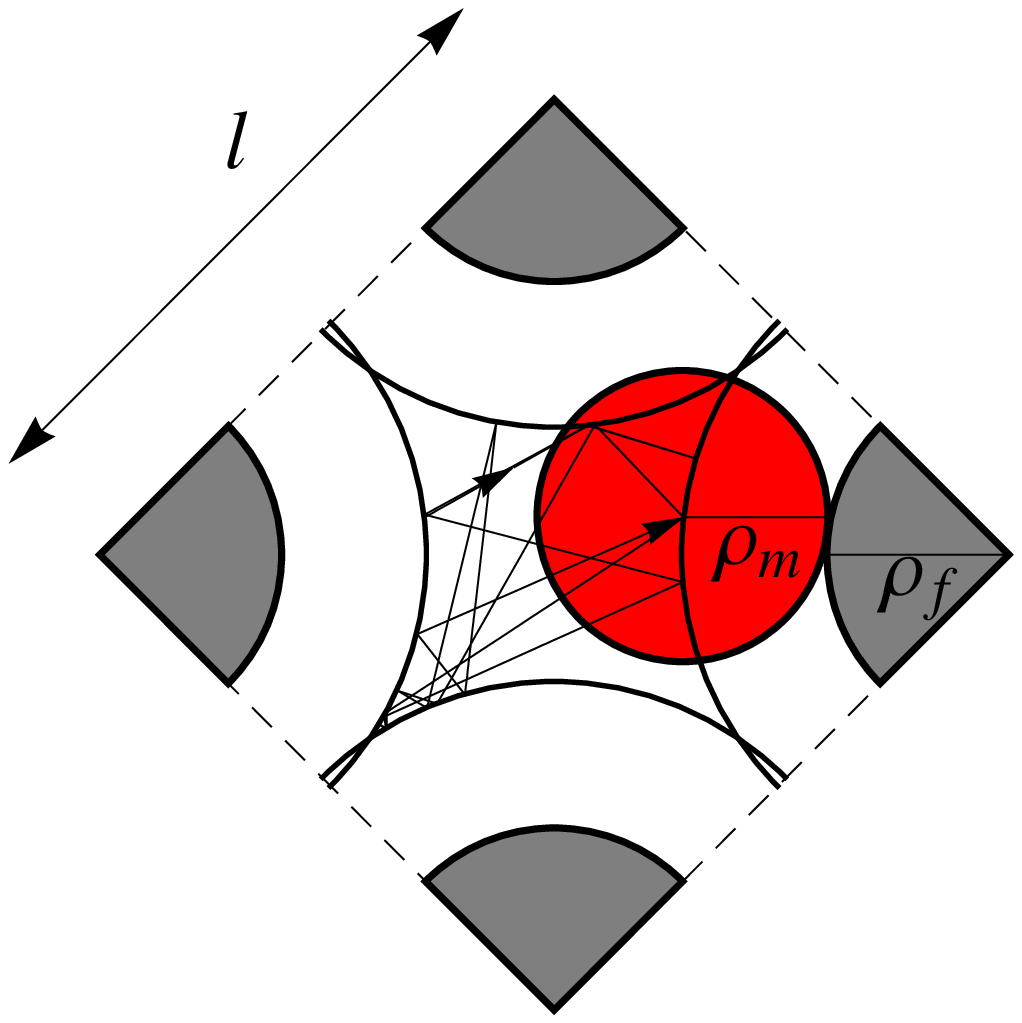}
   }
   \caption{Two equivalent representations of a dispersing billiard table~:
    (left) a point particle moves uniformly inside the domain
    $\mathcal{B}_\rho$ and performs specular collisions with its boundary;
    (right) a disc of radius $\rhom$ moves uniformly inside the domain
    $\mathcal{B}_\rho$ and performs specular collisions with fixed discs of
    radii $\rhof = \rho-\rhom$. The radius $\rhom$ is a free parameter
    which is allowed to take any value between 0 and $\rho$.}
   \label{fig.dispbill}
\end{figure}

As illustrated in figure \ref{fig.dispbill}, the motion of a point particle
in this environment is a limiting case of a class of equivalent dispersing
billiards, whereby the point particle becomes a moving disc with radius
$\rhom$, $0 \leq \rhom \leq \rho$, and bounces off fixed discs of radii
$\rhof = \rho - \rhom$. In all these cases, the motion of the center of the
moving disc is equivalent to that of the point particle in
$\mathcal{B}_\rho$. The border of the domain $\mathcal{B}_\rho$ constitutes
the walls of the billiards.

In the absence of cusps (which occur at $\rho = l/2$), the ergodic
and hyperbolic properties of these billiards are well established
\cite{CM06}. In particular, the long term statistics of the billiard map,
which takes the particle from one collision event to the next, preserves
the measure $\cos\phi\,\ud r\,\ud\phi$, where $\phi$ denotes the angle that the
particle post-collisional velocity makes with respect to the normal vector
to the  boundary. A direct consequence of this invariance is a general
formula which relates the mean free path, $\ell$, to the billiard table
area, $|\mathcal{B}_\rho|$, and perimeter $|\partial \mathcal{B}_\rho|$,
$\ell = \pi|\mathcal{B}_\rho|/|\partial\mathcal{B}_\rho|$.

\begin{figure}[hbt]
   \centerline{\includegraphics[width=.55\textwidth]{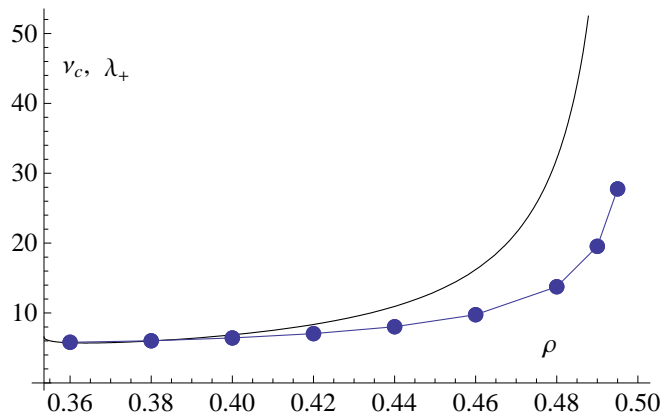}
   }
   \caption{Collision frequency $\nuc$ (solid line) and numerical
     computation of the corresponding
     positive Lyapunov exponent of the flow (dots) of dispersing billiards
     such as shown in figure \ref{fig.dispbill}. Here we took  $\overline{v}
     = 1$ and the square side to be $l = 1/\sqrt{2}$.}
   \label{fig.collfreq}
\end{figure}

The ratio between the speed of the particle, which we denote
$\overline{v}$, and the mean free path gives the wall collision
frequency\footnote{Anticipating the more general definition of the wall
   collision frequency for interacting particle billiard cells, we adopt
   the subscript ``c'' in reference to ``critical'' for reasons to be
   clarified below.}, $\nuc = \overline{v}/\ell$, whose computation is shown
in figure \ref{fig.collfreq}. With this quantity, one can relate the
billiard map iterations to the time-continuous dynamics of the flow. In
particular, the billiard map has two Lyapunov exponents, opposite in signs
and equal in magnitudes, which, multiplied by the wall collision frequency,
correspond to the two non-zero Lyapunov exponents of the flow (there are
two additional zero exponents related to the direction of the flow and
conservation of energy). The results of numerical computations of the
positive one, denoted $\lambda_+$, are shown in figure \ref{fig.collfreq}
for different values of the parameters $\rho$.

Thus let $Q(i,j)$ denote the rhombus of sides $l$ centered
at point
\begin{equation}
(c_{ij},d_{ij}) =
\left\{
\begin{array}{l@{\quad}l}
(\sqrt{2}l i, l j/\sqrt{2}), &j\ \mathrm{even},\\
(\sqrt{2}l i + l/\sqrt{2}, l j/\sqrt{2}), &j\ \mathrm{odd}.\\
\end{array}
\right.
\label{cellcenter}
\end{equation}
The rhombic billiard cell
illustrated in figure \ref{fig.dispbill} becomes a domain centered at point
$(c_{ij},d_{ij})$, defined according to
\begin{equation}
   \mathcal{B}_\rho(i,j) = \Big\{(x, y) \in Q(i,j) \Big|
\delta[(x,y), (c_{ij} + p_k, d_{ij} + q_k)] \ge \rho, k = 1,\dots,d\Big\},
\label{billcell}
\end{equation}
where $\delta[.,.]$ denotes the usual Euclidean distance between two
points, and $(p_k, q_k) = (\pm l/\sqrt{2}, 0), (0, \pm l/\sqrt{2})$ are the
coordinates of the $d \equiv 4$ discs at the corners of the rhombus.

Now consider a number of copies $\{\mathcal{B}_\rho(i,j)\}_{i,j}$ which
tessellate a two-dimensional domain.
We define a {\em lattice billiard} as a collection of billiard cells
\begin{eqnarray}
\mathcal{L}_{\rho,\rhom}(n_1,n_2) &\equiv& \Big\{(x_{ij},y_{ij})\in
\mathcal{B}_\rho(i,j) \Big| \delta[(x_{ij},y_{ij}),(x_{i'j'},y_{i'j'})]
\ge 2\rhom
\nonumber\\
&& \forall\ 1\le i,i'\le n_1,1\le j,j' \le n_2, i\ne i',j\ne j'\Big\}.
\end{eqnarray}
Each individual cell of this billiard table possesses a single moving
particle of radius $\rhom$, $0\le \rhom\le \rho$ and unit mass. All the
moving particles are assumed to have independent initial coordinates within
their respective
cells, with the proviso that no overlap can occur between any pair of
moving particles. The system energy, $E = N \kb T$ (with $N = n_1\times
n_2$, the number of moving particles, $T$ the system temperature,  and
$\kb$ Boltzmann's constant) is constant and assumed to be initially
randomly divided among
the kinetic energies of the moving particles, $E = \sum_{i,j}
\epsilon_{ij}$, $\epsilon_{ij} = mv_{ij}^2/2$, where $v_{ij}$ denotes the
speed of particle $(i,j)$.

Energy exchanges occur when two moving particles located in
neighbouring cells collide. Such events can take place provided the radii of
the moving particles $\rhom$ is large enough compared to $\rho$. Indeed the
value of the critical radius, below which binary collisions do not occur,
is determined by half the separation between the corners of two neighbouring
cells,
\begin{equation}
\rhoc = \sqrt{\rho^2 - \frac{l^2}{4}}.
\label{critcialradius}
\end{equation}

\begin{figure}[htb]
   \centerline{\includegraphics[width=.45\textwidth]{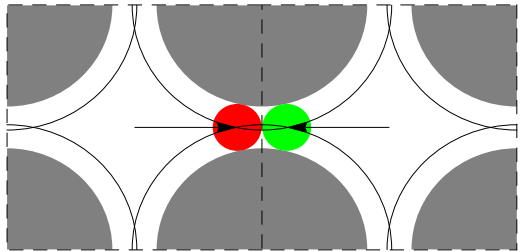}
   }
   \caption{A binary collision event in the critical configuration where
     $\rhom = \rhoc$ would occur only provided the colliding particles
     visit the corresponding corners of their cells simultaneously. The
     value of $\rho$ in this figure is the  same as in  figures
     \ref{fig.dispbill} and \ref{fig.example}.}
   \label{fig.critical}
\end{figure}
For the sake of illustration, the unlikely occurrence of a binary collision
event at the critical radius $\rhom = \rhoc$ is shown in figure
\ref{fig.critical}.

All the collisions are elastic and conserve energy, so that the dynamical
system is Hamiltonian with $2N$ degrees of freedom. Its phase space of
positions and velocities is $4N$-dimensional. Accordingly, the sensitivity
to initial conditions of the dynamics is characterized by $4N$ Lyapunov
exponents, $\{\lambda_i\}_{i=1}^{4N}$, obeying the pairing rule of
symplectic systems, $\lambda_{4N-i+1}=-\lambda_i$, $i=1,\dots,2N$.

Collision events between two moving particles are referred to
as {\em binary collision events} and will be distinguished from {\em 
wall collision
events}, which occur between the moving particles and the walls of their
respective confining cells. The occurrences of the former are characterized
by a {\em binary collision frequency}, $\nub$, and
the latter by a {\em wall collision frequency}, $\nuw$. Both frequencies
depend on the difference $\rhom-\rhoc$, separating the moving particles
radii from the critical radius, equation (\ref{critcialradius}).
By definition of $\rhoc$, the binary collision frequency vanishes
at $\rhom=\rhoc$, $\nub|_{\rhom=\rhoc} = 0$, and, correspondingly,
the wall collision frequency at the critical radius is the collision
frequency of the single-cell billiard, $\nuw|_{\rhom=\rhoc} =
\nuc$. We will assume from now, unless otherwise stated, that the system is
globally isolated and apply periodic boundary conditions at the borders,
thereby identifying $\mathcal{B}_\rho(i + k n_1, j + l n_2)$ with
$\mathcal{B}_\rho(i, j)$ for any $k, l\in \mathbb{Z}$, $1\le i \le n_1$,
$1\le j\le n_2$.

\begin{figure}[hbp]
   \centering
   \includegraphics[width = .53\textwidth]{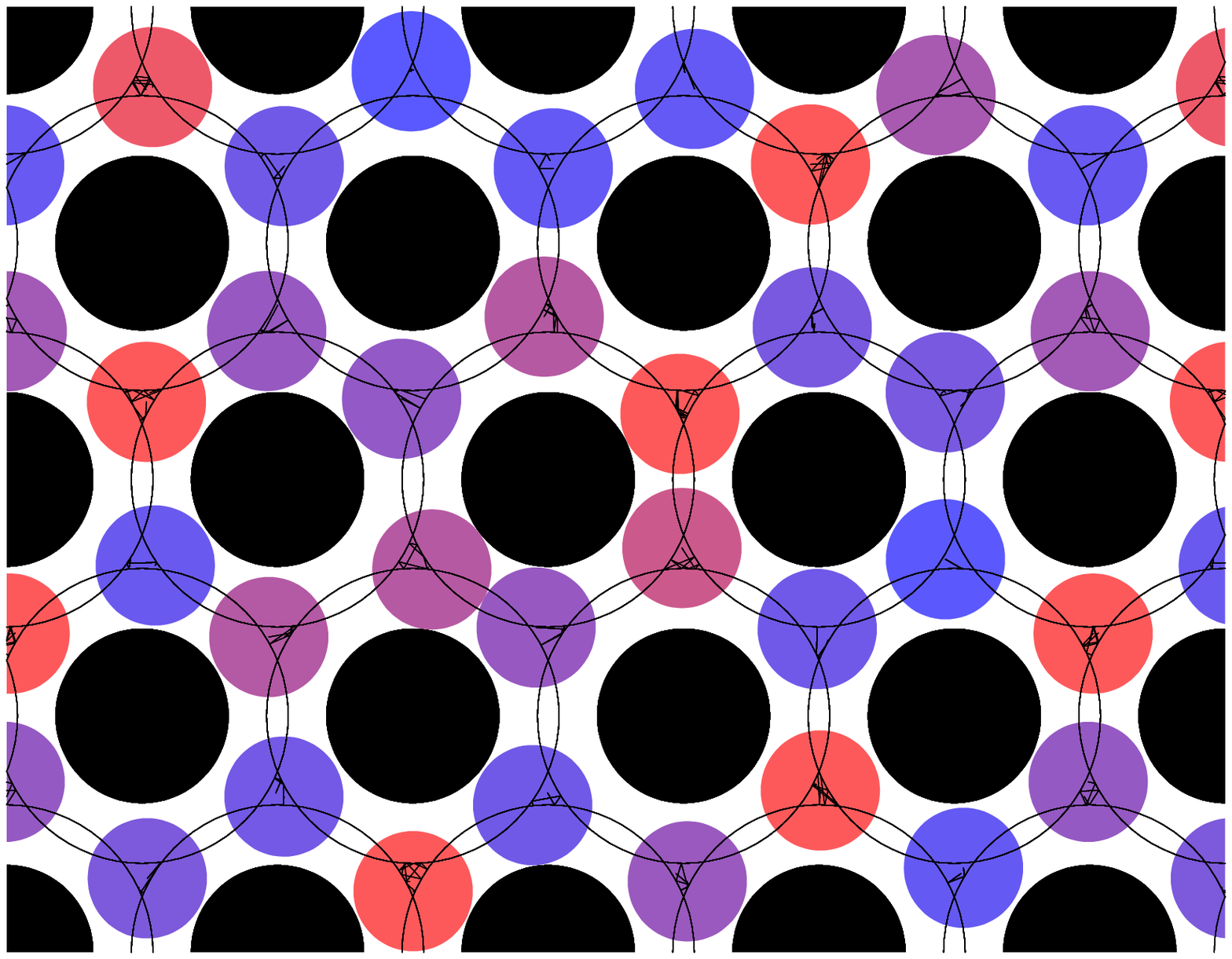}
   \vskip .5cm
   \includegraphics[width = .53\textwidth]{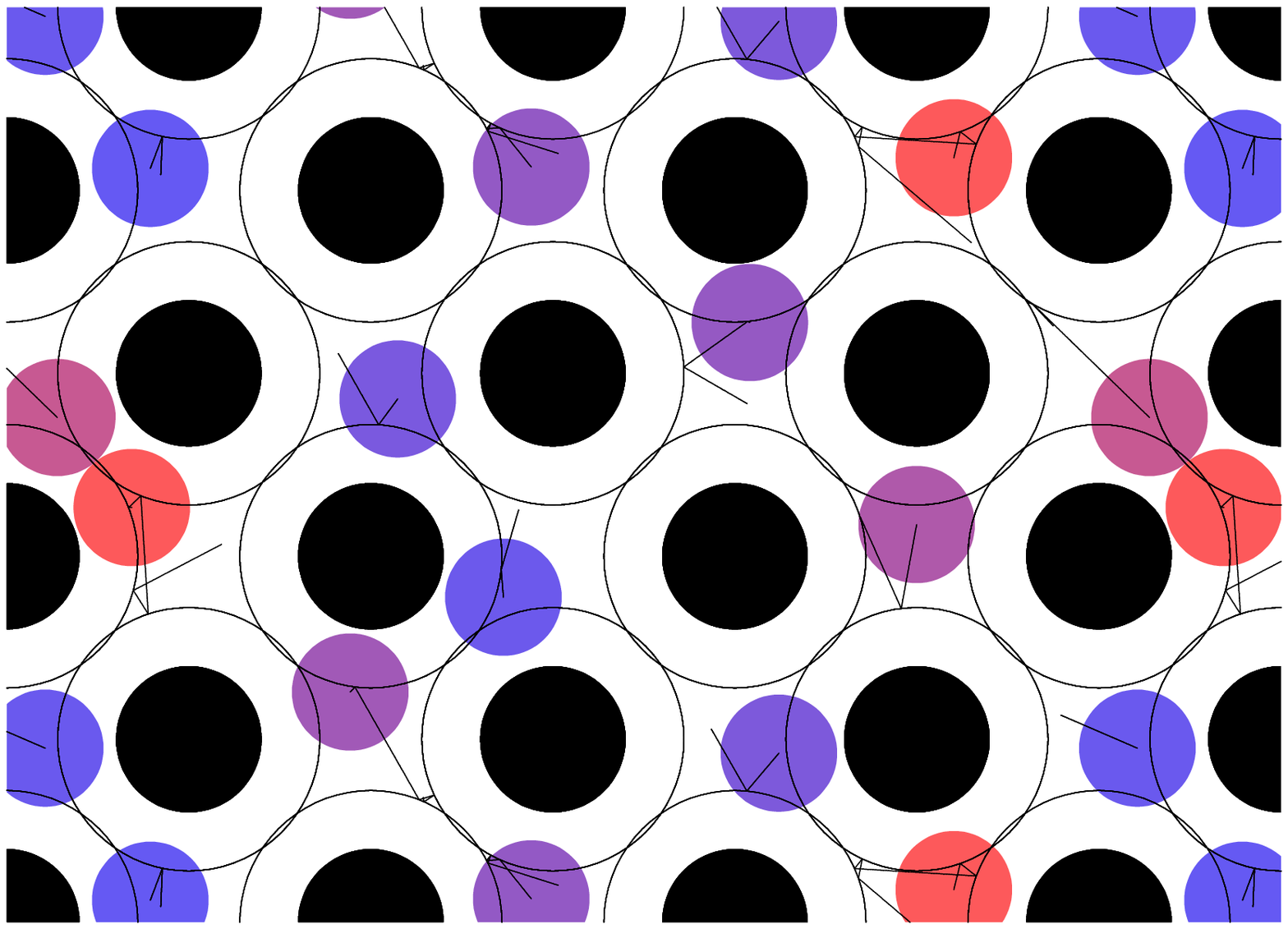}
   \vskip .5cm
   \includegraphics[width = .53\textwidth]{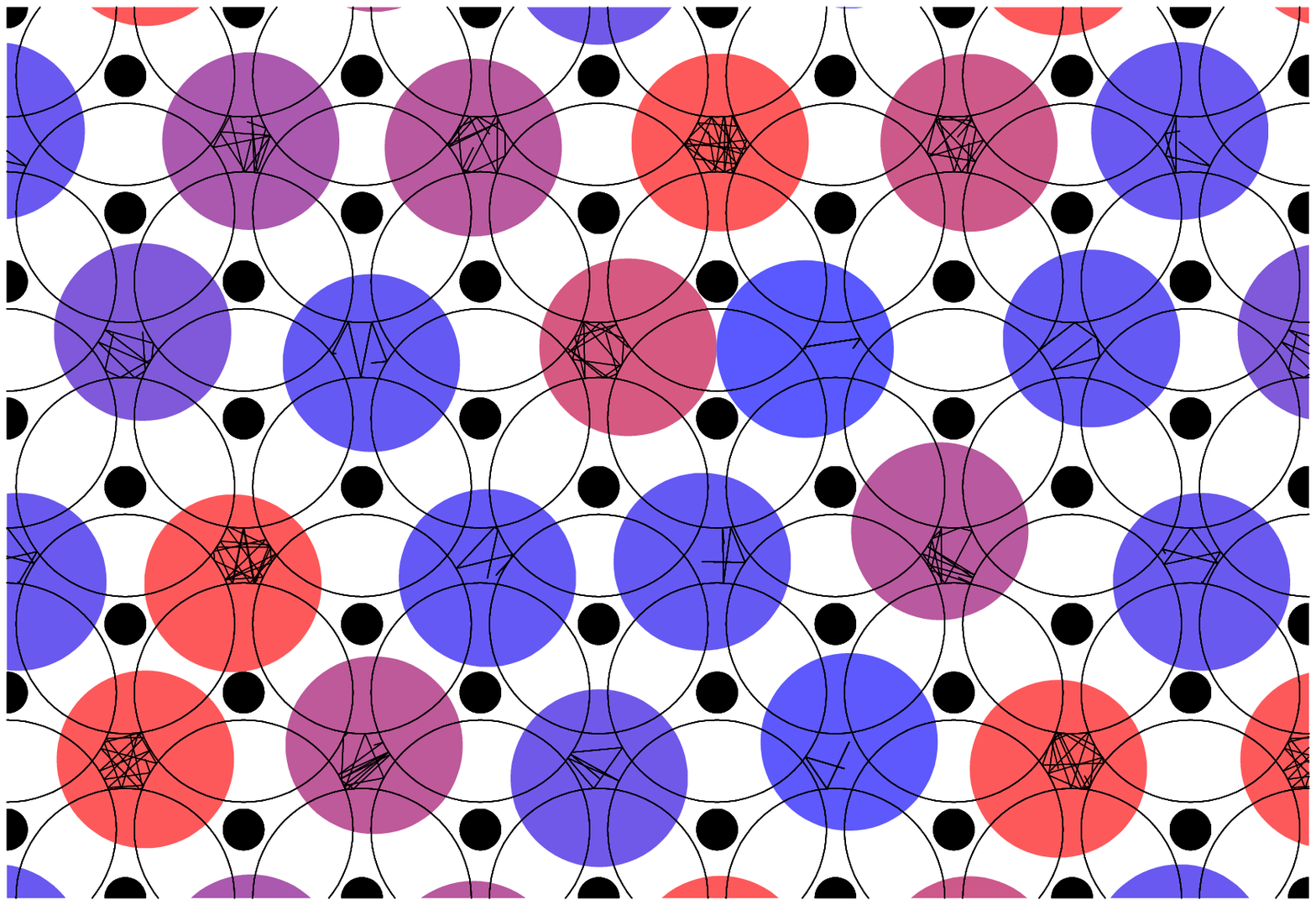}
   \caption{Examples of lattice billiards with  triangular (top), rhombic
     (middle) and hexagonal (bottom) tilings. The coloured particles move among
     an array of
     fixed black discs. The radii of both fixed and mobile discs are chosen
     so that (i) every moving particle is geometrically confined to its own
     billiard cell (identified as the area delimited by the exterior
     intersection of the black circles around the fixed discs), but (ii) can
     nevertheless exchange energy with the moving particles in the
     neighbouring cells through binary collisions. The solid broken lines
     show the trajectories of the moving particles centers about their
     respective cells. The colours are coded according to the particles
     kinetic temperatures (from blue to red with increasing temperature).}
   \label{fig.example}
\end{figure}

Examples of such billiards are displayed
in figure \ref{fig.example}. Obviously the quincunx rhombic lattice
structure, which is generated by the rhombic cells, is but one among
different possible structures. Triangular, upright square, or hexagonal
cells can be used as alternative periodic structures. One might also
cover the plane with random or quasi-crystalline tessellations. The only
relevant assumptions in what follows is that the moving particles must be
confined to their (dispersing) billiard cells and that binary interactions
between neighbouring cells can be turned on and off by tuning the system
parameters.

The two important features of such lattice billiards is that (i) there is
no mass transport across the billiard cells since the moving particles are
confined
to their respective cells, and (ii) energy transport can occur through
binary collision events which take place when the particles of two
neighbouring cells come into contact. In periodic structures such as the
quincunx rhombic lattice, the possibility of such collisions is controlled
by tuning the parameter $\rhom$ about the critical
radius $\rhoc$, keeping $\rho$ fixed.

We can therefore distinguish two separate regimes~:
\begin{itemize}
\item {\bf Insulating billiard cells}: $0\le\rhom<\rhoc$\\
   Absence of interaction between the moving discs. No
   transport process across the individual cells can happen;
\item {\bf Conducting billiard cells}: $\rhoc<\rhom<\rho$\\
   Binary collision events are possible. Energy
   transport across the individual cells takes place.
\end{itemize}
The case $\rhom=\rhoc$ is singular. We will refer to the critical geometry
as the limit $\rhom\stackrel{>}{\to}\rhoc$.

In the insulating regime, there is no interaction among moving particles so
that the billiard cells are decoupled. The moving particles are independent
and their kinetic energies are individually conserved,
resulting in $2 N$ zero Lyapunov exponents. The equilibrium measure in turn
has a product structure and phase-space distributions are locally uniform
with respect to the particles positions and velocity directions. The $N$
positive Lyapunov exponents of the system are all equal to the positive
Lyapunov exponent of the single-cell dispersing billiard,
up to a factor corresponding to the particles speeds, $v_{ij} =
\sqrt{2\epsilon_{ij}/m}$~: ${\lambda_{ij}}_+ = v_{ij}\lambda_+$, where
$\lambda_+$ is the Lyapunov exponent of the single-cell billiard measured
per unit length.

When particles are allowed to interact, on the other hand, local energies
are exchanged through collision events. Thus only the total energy is
conserved in the conducting regime. The ergodicity of such systems of
geometrically confined particles
in interaction was proven by Bunimovich {\em et al.} \cite{BLPS92}. The
resulting dynamical system, whose equilibrium measure is the microcanonical
one (taking into consideration that particles are otherwise uniformly
distributed within their respective cells), enjoys the $K$-property. This
implies ergodicity, mixing, and strong chaotic properties, including the
positivity of the Kolmogorov-Sinai entropy. Two Lyapunov exponents are
zero, one associated to the conservation of energy, the other to the
direction of the flow. The $2(2N-1)$ remaining Lyapunov exponents form
non-vanishing pairs of exponents with opposite signs,
$\lambda_1>\dots>\lambda_{2N-1}>0$, $\lambda_{4N-i+1} = -\lambda_{i}$,
$i = 1, \dots, 2N-1$.

The regime of interest to us is that corresponding to particles interacting
rarely, which is to say, in analogy with a solid, that particles mostly
vibrate inside their cells, ignorant of each other, and only seldom
making collisions with their neighbours, thereby exchanging energy. As
we turn on the interaction and let $\rhom\gtrsim\rhoc$, binary collisions,
though they can occur, will remain unlikely. This is to
say that the binary collision frequency, $\nub$, will, in this regime,
remain small with respect to the wall collision frequency, $\nuw$, which in
the absence of interaction and, in particular, at the critical geometry, we
recall is
equal to the wall collision frequency of the single-cell billiard,
$\nuw|_{\rhom=\rhoc} = \nuc$. When $\rhom\gtrsim\rhoc$, we therefore
expect $\nuw \gg \nub$, as well as $\nuw \simeq \nuc$. In words~: {\em time
   scales separate}. The consequence is that relaxation to local
equilibrium --{\em i.~e.} uniformization of the distribution of the
particles positions and velocity directions at fixed speeds-- occurs
typically much faster than the energy exchange which drives the relaxation
to the global equilibrium. This mechanism justifies resorting to kinetic
theory in order to compute the transport properties of the model.

\section{Kinetic theory \label{sec.kt}}

\subsection{From Liouville's equation to the master equation}

The phase-space probability density is specified by the $N$-particles
distribution function
$p_N(\bi{r}_1,\bi{v}_1,\dots,\bi{r}_N,\bi{v}_N,t)$, where
$\bi{r}_a$ and $\bi{v}_a$, $a = 1,\dots, N$, denote the $a$th
particle position and velocity vectors. The index $a$ stands for the label
$(i,j)$ of the cells defined by equation (\ref{billcell}). For our system,
as is customary for hard sphere dynamics,  this distribution satisfies a
pseudo-Liouville equation \cite{EDHVL69}, which is well defined despite the
singularity of the hard-core interactions. This
equation, which describes the time evolution of $p_N$ is composed of three
types of terms: (i) the advection terms, which account for the displacement
of the moving particles within their respective billiard cells; (ii) the
wall collision terms, which account for the wall collision events, between
the moving particles and the $d$ fixed scattering discs which form the
cells walls; and (iii) the binary collision terms, which account for
binary collision events, between moving particles belonging to neighbouring
billiard cells~:
\begin{equation}
\partial_t p_N = \sum_{a = 1}^N \left[
-\bi{v}_a \cdot \partial_{\bi{r}_a}
+ \sum_{k=1}^d K^{(a,k)}\right] p_N
+ \frac{1}{2}\sum_{a,b=1}^N B^{(a,b)} p_N.
\label{pseudoliouville}
\end{equation}
Each wall collision term involves a single moving particle with index
$a$ and one of the $d$ fixed discs in the corresponding cell, with index
$k$ and position $\bi{R}_k$. Let $\bi{r}_{ak} = \bi{r}_a -
\bi{R}_k$ denote their relative position. Following \cite{RL77}, we have
\begin{eqnarray}
\lefteqn{K^{(a,k)} p_N(\dots,\bi{r}_a, \bi{v}_a,\dots)
=}\nonumber\\
&&\rho \int_{\hat{\bi{e}}\cdot\bi{v}_a>0}
\ud\hat{\bi{e}}(\hat{\bi{e}}\cdot\bi{v}_a)
\Big[\delta(\bi{r}_{ak} - \rho\hat{\bi{e}})
p_N\big(\dots,\bi{r}_a, \bi{v}_a - 2 \hat{\bi{e}}
(\hat{\bi{e}}\cdot\bi{v}_a),\dots\big)
\nonumber\\
&&\hskip 3.5cm -
\delta(\bi{r}_{ak} - \rho\hat{\bi{e}})
p_N\big(\dots,\bi{r}_a, \bi{v}_a,\dots\big)\Big],
\label{wallcollisions}
\end{eqnarray}
where $\hat{\bi{e}}$ denotes the normal unit vector to the fixed disc
$k$ in the cell of particle $a$.

Likewise the binary collision operator, written in terms of the relative
positions $\bi{r}_{ab}$ and velocities $\bi{v}_{ab}$ of particles
$a$ and $b$, and the unit vector $\hat{\bi{e}}_{ab}$ that connects them, is
\begin{eqnarray}
\lefteqn{B^{(a,b)} p_N(\dots,\bi{r}_a, \bi{v}_a,\dots,
\bi{r}_b, \bi{v}_b, \dots)
=}\nonumber\\
&&2 \rhom \int_{\hat{\bi{e}}_{ab}\cdot\bi{v}_{ab}>0}
\ud\hat{\bi{e}}_{ab}(\hat{\bi{e}}_{ab}\cdot\bi{v}_{ab})
\Big[\delta(\bi{r}_{ab} - 2\rhom \hat{\bi{e}}_{ab})
\nonumber\\
&&\times p_N\big(\dots,\bi{r}_a, \bi{v}_a -
\hat{\bi{e}}_{ab}(\hat{\bi{e}}_{ab}\cdot\bi{v}_{ab}), \dots,
\bi{r}_b,  \bi{v}_b +
\hat{\bi{e}}_{ab}(\hat{\bi{e}}_{ab}\cdot\bi{v}_{ab})
,\dots\big) \nonumber\\
&&\hskip 3.5cm - \delta(\bi{r}_{ab} + 2\rhom\hat{\bi{e}}_{ab})
p_N\big(\dots, \bi{r}_a, \bi{v}_a,\dots,\bi{r}_b, \bi{v}_b,
\dots\big) \Big].
\label{binarycollisions}
\end{eqnarray}
We notice that only the terms $B^{(a,b)}$ corresponding to first neighbours
are non-vanishing and contribute to the double sum on the RHS of
equation (\ref{pseudoliouville}).

Provided we have a separation of time scales between wall and binary
collisions, the advection and wall collision terms on the RHS of
equation (\ref{pseudoliouville}) will typically dominate the dynamics on the
short time $\sim 1/\nuw$, which follows every binary collision event, thus
ensuring, thanks to the mixing within individual billiard cells, the
relaxation of the phase-space
distribution $p_N$ to local equilibrium well before the occurrence of the
next binary event, whose time scale is $\sim 1/ \nub$. In other words,
$p_N(\bi{r}_1, \bi{v}_1, \dots, \bi{r}_N, \bi{v}_N,t)$
quickly relaxes to a locally uniform distribution, which depends only on
the local energies, justifying the introduction of
\begin{equation}
\fl
\ple_N(\epsilon_1, \dots, \epsilon_N, t) \equiv
\int \prod_{a=1}^N\ud \bi{r}_a\ud \bi{v}_a
p_N(\bi{r}_1,\bi{v}_1,\dots,\bi{r}_N,\bi{v}_N,t)
\prod_{a=1}^N \delta(\epsilon_a -m v_a^2/2),
\label{ple}
\end{equation}
where $v_a \equiv |\bi{v}_a|$.
On the time scale of binary collision events, this distribution
subsequently relaxes to
the global microcanonical equilibrium distribution. This process
accounts for the transport of energy, and can be characterized by
the master equation \cite{MMN84}
\begin{eqnarray}
\fl \partial_t \ple_N(\epsilon_1, \dots, \epsilon_N, t) =
\frac{1}{2}\sum_{a,b = 1}^N
\int \ud\eta \nonumber\\
\lo \times\Big[W(\epsilon_a + \eta, \epsilon_b - \eta| \epsilon_a,
\epsilon_b)
\ple_N(\dots, \epsilon_a + \eta, \dots,\epsilon_b - \eta, \dots, t)
\nonumber\\
- W(\epsilon_a, \epsilon_b | \epsilon_a - \eta, \epsilon_b + \eta)
  \ple_N(\dots, \epsilon_a, \dots, \epsilon_b, \dots, t)\Big],
\label{masterequation}
\end{eqnarray}
where $W(\epsilon_a, \epsilon_b | \epsilon_a - \eta, \epsilon_b + \eta)$
denotes the probability that an energy $\eta$ be transferred from particle
$a$ to particle $b$ as the result of a binary collision event between them.
This equation is a closure for the local equilibrium distribution $\ple_N$,
obtained from equation (\ref{pseudoliouville})
under the assumption that $\nuw\gg\nub$. The first two terms on the RHS of
equation (\ref{pseudoliouville}) are eliminated because they leave
invariant the local distribution $\prod_{a=1}^N \delta(\epsilon_a -m
v_a^2/2)$. There remain the contributions (\ref{binarycollisions}) from the
binary collisions, which, under the assumption that the local distibutions
are uniform with respect to the positions and velocity directions, yield
the following expression of $W$~:  
\begin{eqnarray}
\fl W(\epsilon_a, \epsilon_b | \epsilon_a - \eta, \epsilon_b + \eta) =
\frac{2 \rhom m^2}{(2\pi)^2|\mathcal{L}_{\rho,\rhom}(2)|}
\int\ud\phi\ud \bi{R}
\int_{\hat{\bi{e}}_{ab}\cdot\bi{v}_{ab} > 0}
\ud \bi{v}_a\ud \bi{v}_b
\label{energytransition}\\
\lo \times \hat{\bi{e}}_{ab}\cdot\bi{v}_{ab}\
\delta\left(\epsilon_a - \frac{m}{2}v_a^2\right)
\delta\left(\epsilon_b - \frac{m}{2}v_b^2\right)
  \delta\left(\eta -
   \frac{m}{2}[(\hat{\bi{e}}_{ab}\cdot\bi{v}_a)^2  -
(\hat{\bi{e}}_{ab}\cdot\bi{v}_b)^2 ]\right),
\nonumber
\end{eqnarray}
where the first integration is performed over the positions of the center
of mass, $\bi{R}\equiv (\bi{r}_a + \bi{r}_b)/2$, between the
two particles $a$ and $b$, given that they are in contact and both located
in their respective cells, and over the angle $\phi$ of the unit vector
connecting $a$ and $b$, $\hat{\bi{e}}_{ab} =  (\cos\phi,\sin\phi)$. The
normalizing factor $|\mathcal{L}_{\rho,\rhom}(2)|$ denotes the $4$-volume
of the billiard corresponding to two neighbouring cells $a$ and $b$, which,
with the assumption that $\rhom\gtrsim\rhoc$, can be
approximated by $|\mathcal{L}_{\rho,\rhom}(2)| \simeq
|\mathcal{B}_\rho|^2$. This substitution amounts to neglecting the
overlap between the two particles; see equation (\ref{twocellarea}) for
a refinement of that approximation. We point out that the position
and velocity integrations in equation (\ref{energytransition}) can be
formally decoupled; in this way, we can prove that the transition rate $W$
is given in terms of Jacobian elliptic functions, see \ref{app.w}. 

\subsection{Geometric factor}

As we show in \ref{app.w}, an important property of the master equation
(\ref{masterequation}) is that the factor which accounts for the geometry of
collision events factorizes from the part of the kernel that accounts for
energy exchanges. Therefore, as $\rhom\to\rhoc$, the critical value of the
radius at which binary collision events become impossible, which is the
regime where the billiard properties are accurately described in terms of
the master equation above, the geometric factor  $\int\ud\phi\ud\bi{R}$
encloses the scaling properties of observables with respect to the
billiard geometry. We now compute this quantity.

\begin{figure*}[thb]
   \centerline{
     \includegraphics[width = .18\textwidth]{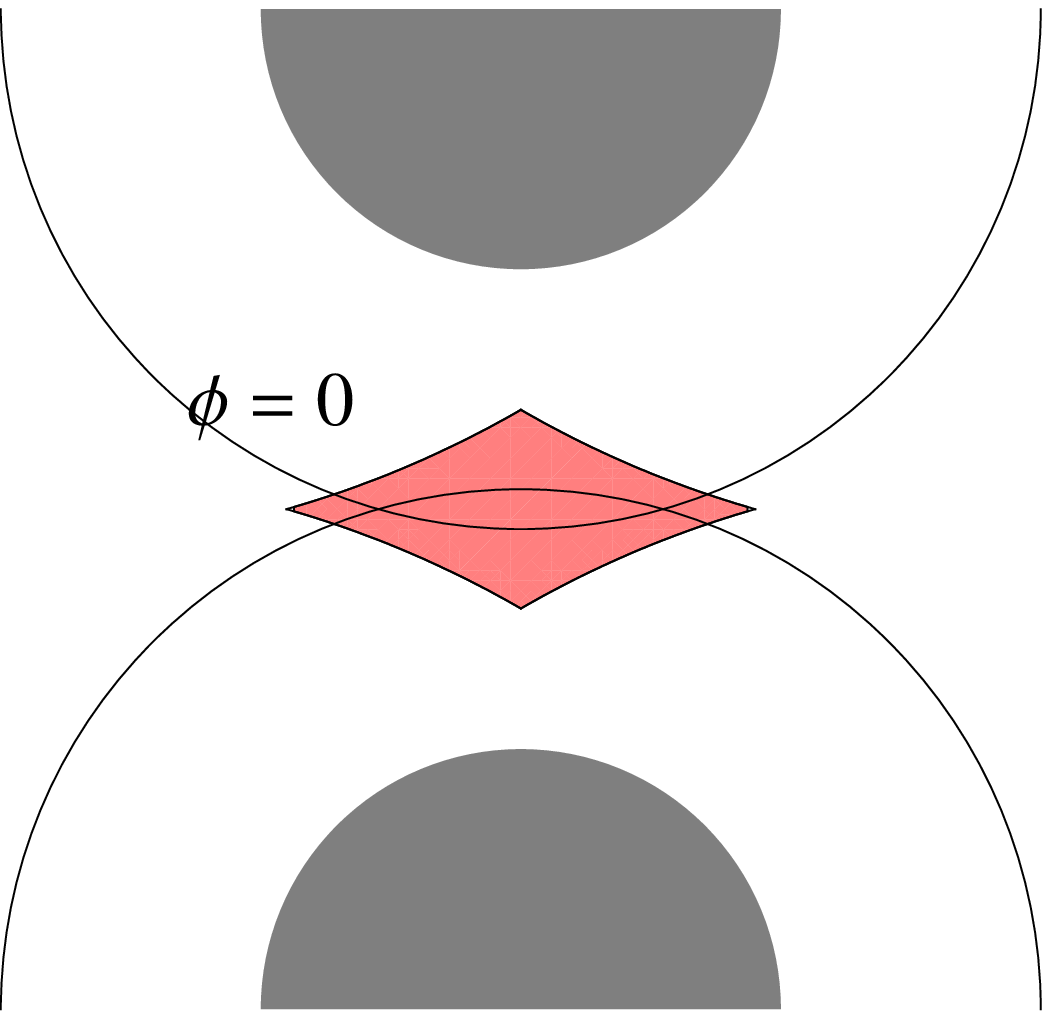}
     \hskip .2cm
     \includegraphics[width = .18\textwidth]{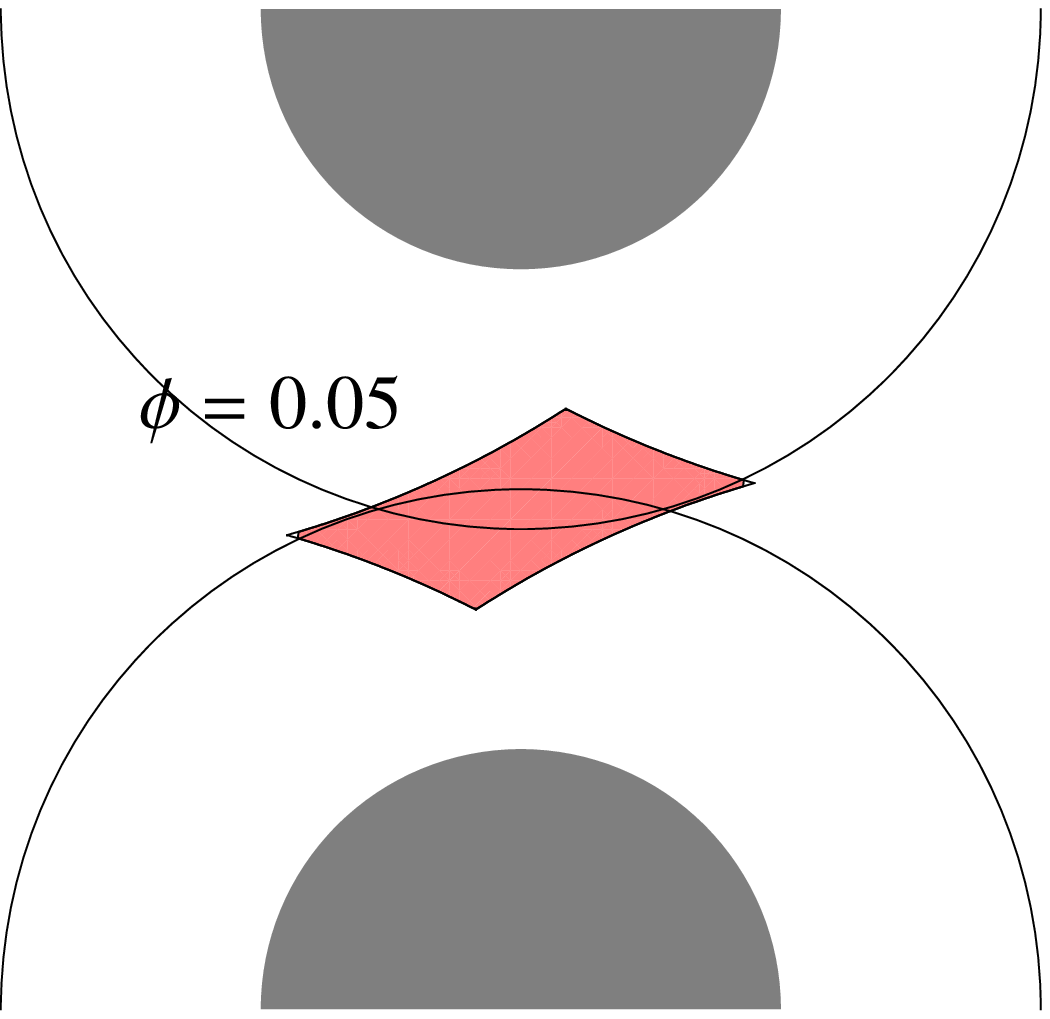}
     \hskip .2cm
     \includegraphics[width = .18\textwidth]{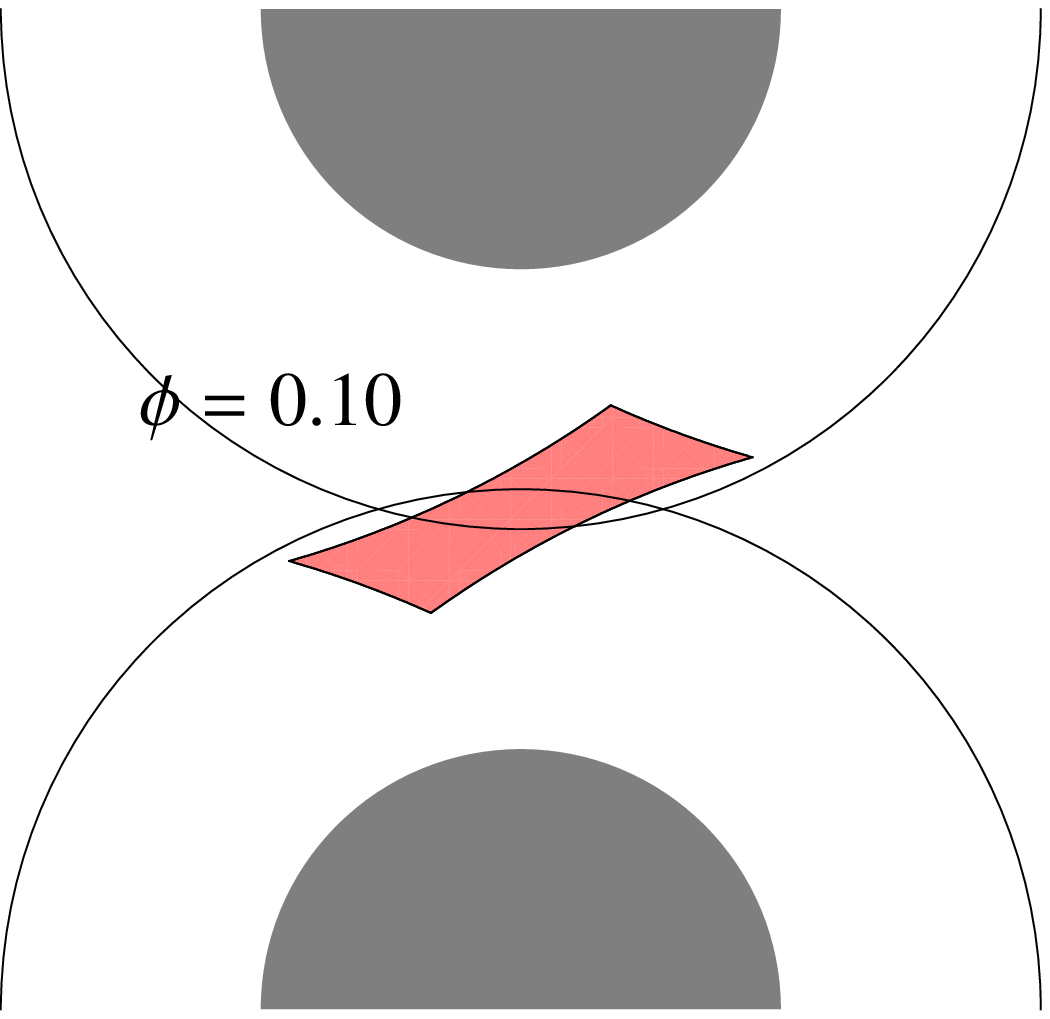}
     \hskip .2cm
     \includegraphics[width = .18\textwidth]{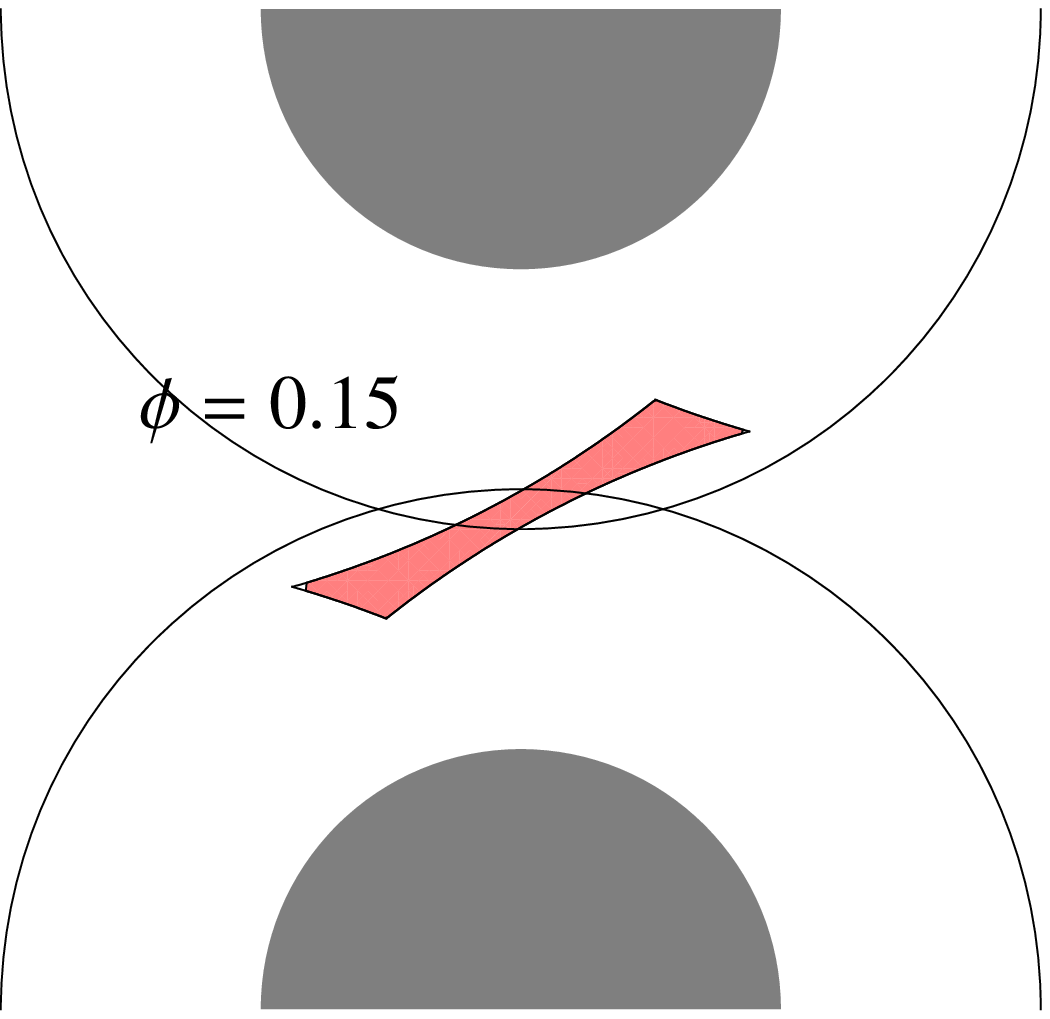}
     \hskip .2cm
     \includegraphics[width = .18\textwidth]{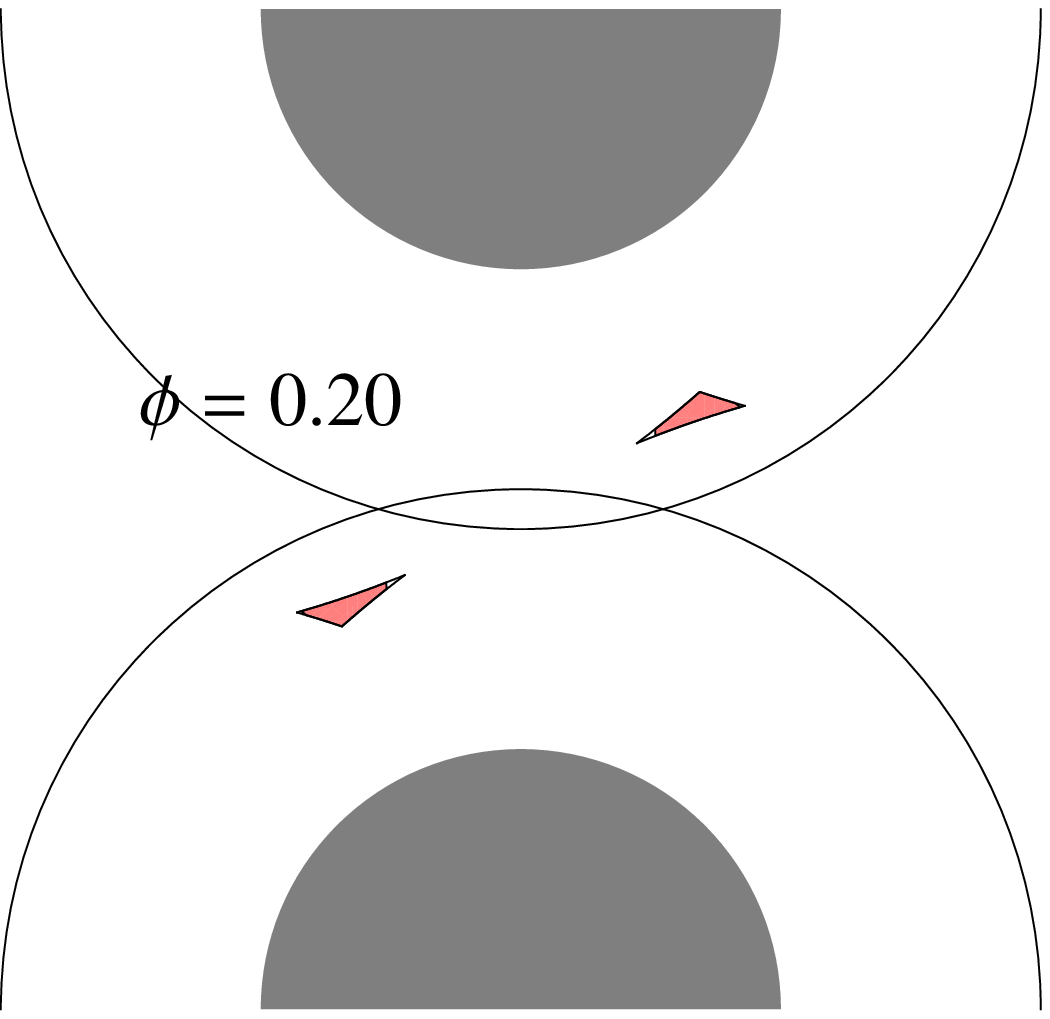}
   }
   \caption{Possible positions of the center of mass $(x,y)$ for different
     values of $\phi$, see equation (\ref{com}), at a binary collision
     event. Here $\rho = 13/25 l$ and $\rhom = 13/50l$.}
   \label{fig.area}
\end{figure*}

A binary collision occurs when particles $a$ and $b$ come to a distance
$2\rhom$ of each other, with $\bi{r}_a\in\mathcal{B}_\rho(a)$ and
$\bi{r}_b\in\mathcal{B}_\rho(b)$. Let $\bi{R} = (x,y)$ be the center
of mass coordinates and $\phi$ be the angle between the particles relative
position and the axis connecting the center of the cells. Taking a
reference frame centered between the cells, we may write
\begin{equation}
\bi{r}_i = \frac{1}{2}(x,y) + \sigma_i \rhom (\cos\phi, \sin\phi),
\label{com}
\end{equation}
where $\sigma_{i} = \pm1$ and $i = a$ or $b$. The integral to be evaluated
is the volume of the triplets $(x,y,\phi)$ about the origin so that
\begin{equation}
\left(\frac{x}{2} + \sigma_i \rhom \cos\phi\right)^2 +
\left(\frac{y}{2} + \sigma_i \rhom \sin\phi \pm \frac{l}{2}\right)^2
\ge \rho^2.
\end{equation}
As illustrated in figure \ref{fig.area}, for different orientations $\phi$
of the vector connecting the two particles, this is a region bounded by
four arc-circles, which we denote by
\begin{equation}
y_{\sigma,\tau}(x) \equiv -2 \sigma \rhom \sin\phi
- \tau l +\tau \sqrt{4\rho^2 - \left(x + 2 \sigma \rhom \cos\phi\right)^2},
\label{curvey}
\end{equation}
where $\sigma, \tau = \pm 1$.

As seen from figure \ref{fig.area}, the area is connected for
$-\phi_\mathrm{T}\leq\phi\leq\phi_\mathrm{T}$, where $\phi_\mathrm{T}$ is
the angle $\phi$ at which opposite arcs intersect,
\begin{equation}
\phi_\mathrm{T} =
\arcsin \frac{\rhom^2 - \rhoc^2}{l\rhom}.
\label{phit}
\end{equation}
Beyond that value, the area splits into two triangular areas. These areas
shrink to zero at the angle $\phi$ given by
\begin{equation}
\phi_\mathrm{M} = \arccos\frac{
\rhom \rhoc + l/2\sqrt{\rho^2 - \rhom^2}}
{\rho^2}.
\label{phim}
\end{equation}
Let $-\phi_\mathrm{M} \leq \phi \leq \phi_\mathrm{M}$. We denote by $x_1 <
x_2 < x_3 < x_4$ the four corners of the rectangular domain,
\begin{equation}
   \begin{array}{lcl}
     x_1 &=& -x_4 = -2(\rhom\cos\phi - \rhoc),\\
     x_2 &=& -x_3 = -2 \sqrt{\rho^2 - \rhom^2} \sin\phi.
   \end{array}
   \label{pointsx}
\end{equation}
For $\phi\ge\phi_\mathrm{T}$, the points at which the opposite arcs intersect
are given by
\begin{equation}
x_i = \pm (l - 2 \rhom \sin \phi) \left[
\frac{-l^2 + 4 l\rhom \sin \phi + 4 (\rho^2 - \rhom^2)}
{l^2 - 4 l\rhom \sin \phi + 4 \rhom^2}\right]^{1/2}.
\label{pointxc}
\end{equation}
Combining equations (\ref{curvey})-(\ref{pointxc}) together, we can make use
of the symmetry $\phi\to-\phi$ and write the integral to be computed as
\begin{equation}
\alpha(\rho, \rhom) \equiv
\int\ud\phi\ud\bi{R}
= 2\left[\int_0^{\phi_\mathrm{M}} A_1(\phi)\ud \phi
+ \int_{\phi_\mathrm{T}}^{\phi_\mathrm{M}} A_2(\phi)\ud \phi
\right],
\label{area}
\end{equation}
where
\begin{equation}
\fl A_1(\phi) = \int_{x_1}^{x_3} y_{+1,-1}(x) \ud x +
\int_{x_3}^{x_4} y_{-1,-1}(x) \ud x
- \int_{x_1}^{x_2} y_{+1,+1}(x) \ud x
- \int_{x_2}^{x_4} y_{-1,+1}(x) \ud x,
\label{A1}
\end{equation}
which is the area bounded by the four arcs $y_{\sigma,\tau}$, and
\begin{equation}
A_2(\phi) = \int_{-x_i}^{x_i} [y_{-1,+1}(x)  -
y_{+1,-1}(x)] \ud x,
\label{A2}
\end{equation}
is the area of the overlapping opposite arcs $y_{-1,+1}$ and $y_{+1,-1}$,
which occurs when $\phi_\mathrm{T}\leq\phi\leq\phi_\mathrm{M}$ [it gives a
negative contribution to $A_1(\phi)$].

\begin{figure}[hbt]
   \centerline{
     \includegraphics[width = .55\textwidth]{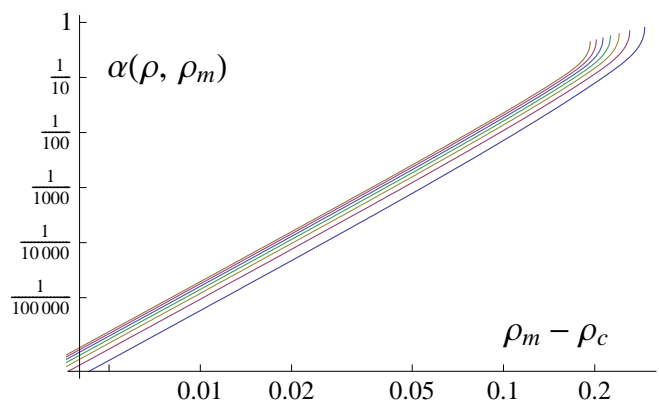}
   }
   \caption{Area of binary collisions $\alpha(\rho, \rhom)$ versus
     $\rhom-\rhoc$ for seven values of $\rho$ ranging from $9/25$ to
     $42/100$ ($l = 1/\sqrt{2}$).}
   \label{fig.areaint}
\end{figure}

The computation of these expressions is easily performed numerically, and
the result shown in figure \ref{fig.areaint}.

Near the critical geometry, we expand the quantity (\ref{area}) in powers
of the difference $\rhom-\rhoc$,
\begin{equation}
   \alpha(\rho,\rhom\big) = \sum_{n=1}^\infty c_n (\rhom-\rhoc)^n.
   \label{alphaexp}
\end{equation}
The first two coefficients vanish, so that the leading term corresponds to
$n=3$. The first few coefficients are derived in \ref{app.coeff}
and given by~:
\begin{equation}
\eqalign{
c_1 = c_2 = 0,\\
c_3 = \frac{128 \rhoc}{3 l^2},\\
c_4 = \frac{256 \rhoc^2}{3 l^4},\\
c_5 = \frac{16}{l^2}\left(\frac{1}{3\rhoc} + \frac{8\rhoc}{5 l^2}
+ \frac{16\rhoc^3}{l^4} + \frac{16\rho^4}{5 l^4 \rhoc}\right).
}
\label{coefficients}
\end{equation}

We further note that the computation of the two-cell 4-volume,
$|\mathcal{L}_{\rho,\rhom}(2)|$, which, as noticed above, is approximated
by the square of the single cell area $|\mathcal{B}_\rho|^2$,
can be improved using equation (\ref{alphaexp}). Indeed
it is easily seen that $\ud|\mathcal{L}_{\rho,\rhom}(2)|/\ud \rhom = -2
\alpha(\rho,\rhom)$, which implies
\begin{equation}
|\mathcal{L}_{\rho,\rhom}(2)| = |\mathcal{B}_\rho|^2 - 2 \sum_{n=4}^\infty
\frac{c_{n-1}}{n}(\rhom - \rhoc)^n.
\label{twocellarea}
\end{equation}
It is immediate to check that corrections $|\mathcal{L}_{\rho,\rhom}(2)|
- |\mathcal{B}_\rho|^2 = \mathcal{O}(\rhom - \rhoc)^4$.

\subsection{Binary collision frequency}

Having computed the transition rates of the master equation with respect
to the billiard geometry, we now turn to the computation of
observables. The first quantity of interest, which can be readily computed
from equation (\ref{energytransition}), is the binary collision frequency 
$\nub$. This is an equilibrium quantity which, in the (global)
microcanonical ensemble with energy $E = \epsilon_1 + \dots + \epsilon_N$,
involves the two-particle energy distribution,
\begin{equation}
   \pmic_2(\epsilon_a, \epsilon_b) = \frac{(N-1)(N-2)}{E^2}
   \left(1 - \frac{\epsilon_a  + \epsilon_b}{E}\right)^{N-3},
\label{pmic2}
\end{equation}
and can be written as
\begin{equation}
\nub = \int \ud\epsilon_a \, \ud\epsilon_b \, \ud \eta \,
W(\epsilon_a, \epsilon_b | \epsilon_a - \eta, \epsilon_b + \eta)
\pmic_2(\epsilon_a, \epsilon_b).
\label{nubKT}
\end{equation}
Taking the large $N$ limit and letting $E = N \kb T \equiv N/\beta$, we can
write $
\pmic_2(\epsilon_a, \epsilon_b) \simeq \beta^2\exp[-\beta(\epsilon_a  +
\epsilon_b)][1+\mathcal{O}(N^{-1})]$.
Substituting this expression into the above equation and
inserting the expression of $W$ from equation (\ref{energytransition}), we
obtain, after some calculations,
\begin{equation}
\nub \simeq \sqrt{\frac{\kb T}{\pi m}}\frac{2\rhom}{|\mathcal{B}_\rho|^2}
\left(\int\ud\phi\ud\bi{R}\right)
[1 + \mathcal{O}(N^{-1})].
\label{nubresult}
\end{equation}
This expression involves the geometric factor (\ref{area}) in the first
bracket. The leading term in the second bracket is the canonical expression
of the binary collision frequency. The second term is a positive finite $N$
correction, which is useful in that it shows that the binary collision
frequency decreases to its asymptotic value as $N\to\infty$. 

\subsection{Rescaled master equation \label{subsec.rescaling}}

The binary collision frequency (\ref{nubresult}) defines a natural
dimensionless time scale for the stochastic process described by the master
equation (\ref{masterequation}) with transition rates
(\ref{energytransition}). The master equation can thus be converted to 
dimensionless form by rescaling the energies by a reference thermal energy
and time by the corresponding asymptotic ($N\to\infty$) value of the binary
collision frequency (\ref{nubresult}).

Introducing the variables
\begin{eqnarray}
e_a &\equiv& \frac{\epsilon_a}{\kb T}, \\
h &\equiv& \frac{\eta}{\kb T}, \\
\tau &\equiv& 
\nub t,
\end{eqnarray}
equation (\ref{masterequation}) becomes
\begin{eqnarray}
\fl \partial_\tau \pple_N(e_1, \dots, e_N, \tau) =
\frac{1}{2}\sum_{a,b = 1}^N
\int \ud h \nonumber\\
\lo \times\Big[w(e_a + h, e_b -h| e_a,
e_b) \; \pple_N(\dots, e_a + h, \dots,e_b -h, \dots, \tau)
\nonumber\\
- w(e_a, e_b | e_a - h, e_b + h) \;
  \pple_N(\dots, e_a, \dots, e_b, \dots, \tau)\Big],
\label{rescaledmasterequation}
\end{eqnarray}
with the transition rates
\begin{eqnarray}
\fl w(e_a, e_b | e_a - h, e_b + h) &=&
\sqrt{\frac{2}{\pi^3}} \int_{x_{1\Vert}-x_{2\Vert}>0} \ud x_{1\Vert} 
\, \ud x_{1\bot} \, \ud x_{2\Vert} \, \ud x_{2\bot}
\, 
(x_{1\Vert}-x_{2\Vert}) \\ && \qquad \times 
\delta(e_a-x_{1\Vert}^2-x_{1\bot}^2)
\, 
\delta(e_b-x_{2\Vert}^2-x_{2\bot}^2)
\, 
\delta(h-x_{1\Vert}^2+x_{2\Vert}^2)
 \nonumber
\end{eqnarray}
This master equation shows that all the properties of heat conduction
are rescaled by the binary collision frequency and temperature
in its limit of validity where the collision frequency vanishes.
In particular, this shows that the coefficient of heat conduction
is proportional to the binary collision frequency in this limit, 
as explained in the next subsection.

\subsection{Thermal conductivity}

Starting from the master equation (\ref{masterequation}), we derive an
equation for the evolution of the kinetic energy of each moving
particle, $\langle \epsilon_a \rangle \equiv \kb T_a$, which defines 
the local temperature, 
where $\langle.\rangle$ denotes an average with respect to the energy
distributions. 
By the structure
of equation (\ref{masterequation}), such an  equation can be expressed in
terms of the transfer of energy due to the binary collisions between
neighbouring cells. The time evolution of the average local kinetic 
energy is given by
\begin{equation}
\partial_t \langle \epsilon_a 
\rangle =  - \sum_b  \langle J_{a,b} (\epsilon_a, \epsilon_b)\rangle ,
\end{equation}
with the 
energy flux defined as
\begin{equation}
\eqalign{
  J_{a,b}(\epsilon_a, \epsilon_b) \equiv \int \ud\eta 
  \, \eta \, W(\epsilon_a, \epsilon_b | \epsilon_a - \eta, \epsilon_b + 
  \eta),\\
  \phantom{J_{a,b}(\epsilon_a, \epsilon_b)} = - \int \ud\eta \, \eta\,
  W(\epsilon_a+\eta, \epsilon_b -\eta | \epsilon_a , \epsilon_b ).
}
\end{equation}
This expresses the local conservation of energy.

Over long time scales, the probability distribution becomes 
controlled by this local conservation of energy, the slowest variables
being the local kinetic energies $\langle \epsilon_a \rangle$ or,
equivalently, the local temperatures $T_a$ as defined above. This holds
even though statistical  correlations develop between the local energies in
the probability  distributions. These statistical correlations are well
known for transport processes ruled by master equations such as
Eq.~(\ref{masterequation}) \cite{MMN84,S91} and are observed in the present
system as well. 

To be specific, we consider a one-dimensional chain, extending along the
$x$-axis, formed with a succession of pairs of rhombic billiard cells
arranged in quincunx, similarly to the middle panel of figure
\ref{fig.example}, except the vertical height is here only one unit of
length. The unit of horizontal and vertical lengths is thus $l \sqrt{2}$
and there are two cells per each unit of length. Similar results hold for
different choices of geometry modulo straightforward adaptations.

We imagine that the system is in a non-equilibrium state, 
with a 
small temperature difference $\delta T$ about
an average temperature $T$ between neighbouring cells, and consider the
average heat transfered from cell $a$ at inverse temperature $\beta_a = \beta +
\delta\beta/2$ to cell $b$  at inverse temperature $\beta_b = \beta -
\delta\beta/2$, $\delta\beta = -\delta T/(\kb T^2)$,
both cells being assumed to be in thermal equilibrium at their respective
temperatures. The statistical correlations we observe in the present 
system are of the order of $\delta\beta^2$, as it is the case in other 
systems \cite{MMN84}. In the non-equilibrium state, these statistical
correlations are controlled in the long-time limit by the local
temperatures. Since the process is here ruled by a Markovian master  
equation in the limit of small binary collision frequency, we get the
equation of heat for the temperature 
\begin{equation}
\partial_t T(x,t) = \partial_x [\kappa \partial_x T(x,t)],
\label{fourierlaw}
\end{equation}
where the local temperature is here written as $T(x,t) =  \langle
\epsilon_{ij} \rangle/\kb$, $x = \sqrt{2} l (i+j/2)$, $i=1,\dots,N/2$, $j =
0,1$ [see equation (\ref{cellcenter})].

According to the rescaling property of the master equation discussed in 
section \ref{subsec.rescaling}, the heat conductivity is proportional to
the binary collision frequency \footnote{We notice that the heat capacity
  per particle is equal to $c_V = (1/N) \partial E/\partial T = \kb$, so
  that the thermal conductivity is also equal to the thermal diffusivity in
  units where $\kb = 1$. 
}~:
\begin{equation}
\frac{\kappa}{l^2} = A \, \nub ,
\label{kappanub}
\end{equation}
with a dimensionless constant $A$.

An analytical estimation can be obtained by transforming the master
equation into a hierarchy of equations for all the moments of the
probability distribution: $\langle \epsilon_a\rangle$, $\langle
\epsilon_a\epsilon_b\rangle$, $\langle
\epsilon_a\epsilon_b\epsilon_c\rangle$,... The evolution equations of these
moments are coupled. 

Truncating the hierarchy at the equations for the averages $\langle
\epsilon_a\rangle$, we get the approximate heat conductivity~:
\begin{eqnarray}
\frac{\kappa}{l^2} &\simeq \frac{\beta^4}{2}\int \ud\epsilon_a \ud\epsilon_b
\ud\eta \ \eta(\epsilon_b - \epsilon_a) W(\epsilon_a, \epsilon_b|\epsilon_a
- \eta, \epsilon_b + \eta)
  \exp[-\beta(\epsilon_a + \epsilon_b)],\nonumber\\
&= \sqrt{\frac{\kb T}{\pi m}}\frac{2\rhom}{|\mathcal{B}_\rho|^2}
\left(\int\ud\phi\ud\bi{R}\right).
\label{heatconductivity}
\end{eqnarray}
So that $A = 1$ in this approximation. The same result holds if we 
include the equations for the moments $\langle \epsilon_a
\epsilon_b\rangle$ with $\vert a-b\vert \leq 1$. 

Though the approximate result (\ref{heatconductivity}) above does not rule
out possible corrections to $A=1$, we argue that $A=1$ is an exact property
of the master equation 
(\ref{masterequation}) in the infinite system limit. This claim is borne
out by extensive studies of the 
stochastic process described by equation (\ref{masterequation}) that will be
reported in a separate publication \cite{GG08b}. The focus is here on the
billiard systems whose conductivity may however bear corrections to this
identity, due we believe to lack of sufficient separation of time scales
between wall and binary collision events. In the following, the results of
numerical computations are presented which support these claims.

\section{Numerical results \label{sec.nr}}

The above formulae for the binary collision frequency,
equation (\ref{nubresult}), and the thermal conductivity,
equation (\ref{heatconductivity}), together with the expressions of the
geometric factors, equations (\ref{coefficients}) and (\ref{twocellarea}),
provide a detailed picture of the mechanism
which governs the transport of heat in our model. Numerical computations
of these quantities further add to this picture and provide strong evidence
of the validity of our theoretical approach.

\subsection{Binary and wall collision frequencies}

For the sake of computing the binary collision frequency, we simulate the
quasi-one dimensional channel of $N$ cells with rhombic shapes and
apply periodic boundary conditions at the horizontal ends of the
channel. Thus each cell has two neighbouring cells, left and right, and
interactions between any two neighbouring cells can occur through both top
and bottom corners \footnote{We mention that this set-up allows for
   re-collision between two particles (under stringent conditions), due to
   the vertical periodic boundary conditions.  This conflicts with the
   assumptions of \cite{BLPS92}, but does not seem to affect the results as
   far as numerics are concerned.}.

\begin{figure}[htp]
   \centering
   \includegraphics[width = .55\textwidth]{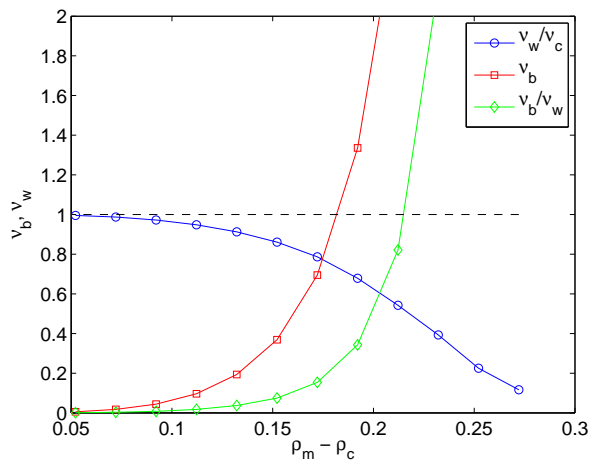}
   \caption{Wall and binary collision frequencies $\nuw$ and $\nub$
     versus $\rhom - \rhoc$ in a one-dimensional channel of $N=10$ rhombic
     cells with $\rho = 9/25$ and lattice spacing $l=1/\sqrt{2}$.}
   \label{fig.nub}
\end{figure}

In figure \ref{fig.nub}, we compare the computations of $\nub$ to $\nuw$
for a system of $N=10$ cells at unit temperature. The parameters are taken
to be $\rho = 9/25$ and $\rhom = 3/25, \dots,
17/50$ by steps of $1/50$. Both collision frequencies are computed in the
units of the microcanonical average velocity,  $\overline{v}_\mathrm{N}
\equiv 2^N\sqrt{N/2} (N-1)!/(2N-1)!!$, which, as $N\to\infty$, converges
to the canonical average velocity $\overline{v}_\infty = \sqrt{\pi/2}$. The
wall collision frequency is compared to the collision frequency of the
isolated cells $\nuc$, itself measured in the units of the single particle
velocity. As expected, $\nub/\nuw\ll1$ and $\nuw \simeq
\nuc$ for $\rhom\gtrsim\rhoc$. The crossover $\nub\simeq\nuw$ occurs at
$\rhom \simeq 11/50$ for this value of $\rho$.

\subsection{Thermal conductivity}

\subsubsection{Heat flux.}

The thermal conductivity can be obtained by
computing the heat flux in a nonequilibrium stationary state. Such
stationary states occur when the two ends of the channel are put in contact
with heat baths at separate temperatures, say $T_-$ and $T_+$, $T_-<T_+$.

Let a system of $N$ cells be in contact with two thermostated cells at
respective temperatures $T_\pm$, and let these cell indices be $n=\pm
(N+1)/2$ (we take $N$ odd for the sake of definiteness). Provided the
difference between the bath temperatures is small, $T_+ - T_- \ll (T_+ +
T_-)/2$, a linear gradient of temperature establishes through the system,
with local temperatures
\begin{equation}
T_n = \frac{1}{2}(T_+ + T_-) + \frac{n}{N+1}(T_+ - T_-).
\label{noneqTn}
\end{equation}
Under these conditions, the nonequilibrium stationary state is expected to
be locally well approximated by a canonical equilibrium at temperature
$T_n$.

This can be checked numerically. In fact the local thermal equilibrium is
verified (by comparing the moments of $\epsilon_n$ to their Gaussian
expectation values) under the weaker property of small local
temperature gradients, {\em i.~e.} $(T_+ - T_-)/N \ll (T_+ + T_-)/2$, for
which the temperature profile is generally not linear since the thermal
conductivity depends on the temperature. Indeed, since $\kappa\propto
T^{1/2}$, we expect in that case, according to Fourier's law, the profile
\begin{equation}
T_n = \left[\frac{1}{2}(T_-^{3/2} + T_+^{3/2}) + \frac{n}{N+1}
(T_+^{3/2} - T_-^{3/2})\right]^{2/3}.
\label{noneqTn_nl}
\end{equation}

The thermal conductivity can therefore be computed from the heat exchanges
of the chain in contact with the two cells at the ends of the chain, 
respectively thermalized at
temperatures $T\pm\delta T/2$, $\delta T\ll T$. The thermalization of the end
cells is achieved by randomizing the velocities of the two particles at
every collision they make with their cells walls, according to the usual
thermalization procedure of particles colliding with thermalized
walls \cite{TTKB98}. 

First, we consider a chain containing a single 
cell in order to
test the validity of the master equation.
In this 
case, the procedure amounts to simulating a single particle
confined to its cell and performing random collisions with stochastic
particles which penetrate the cell corners according to the statistics of
binary collisions. For a chain with a single cell, the heat conductance is
given by Eq. (\ref{heatconductivity}) with $A=1$, 
as there are no correlations with the stochastic particles. 

\begin{figure}[thp]
   \centering
   \includegraphics[width = .45\textwidth]{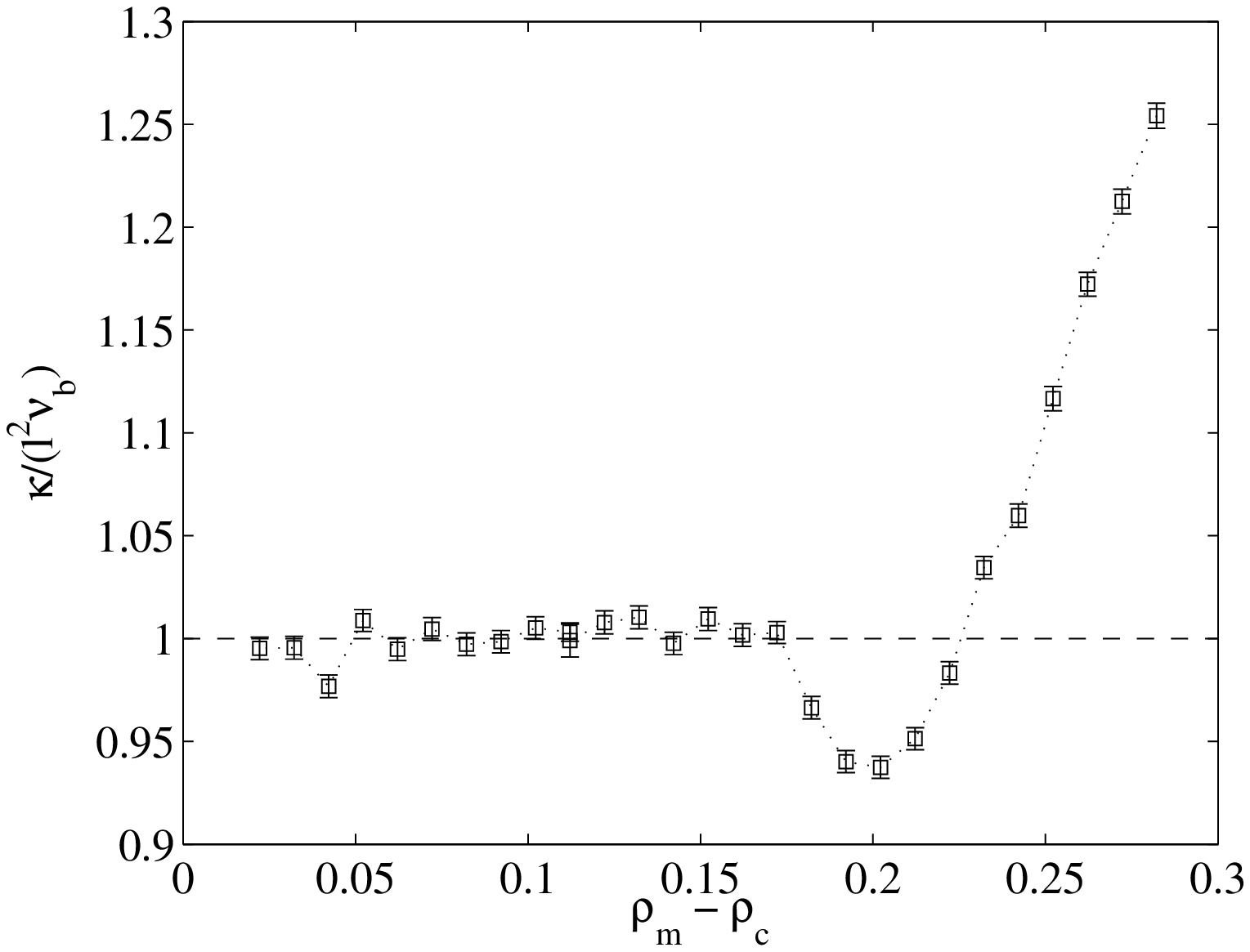}
   \hskip .5cm
   \includegraphics[width = .45\textwidth]{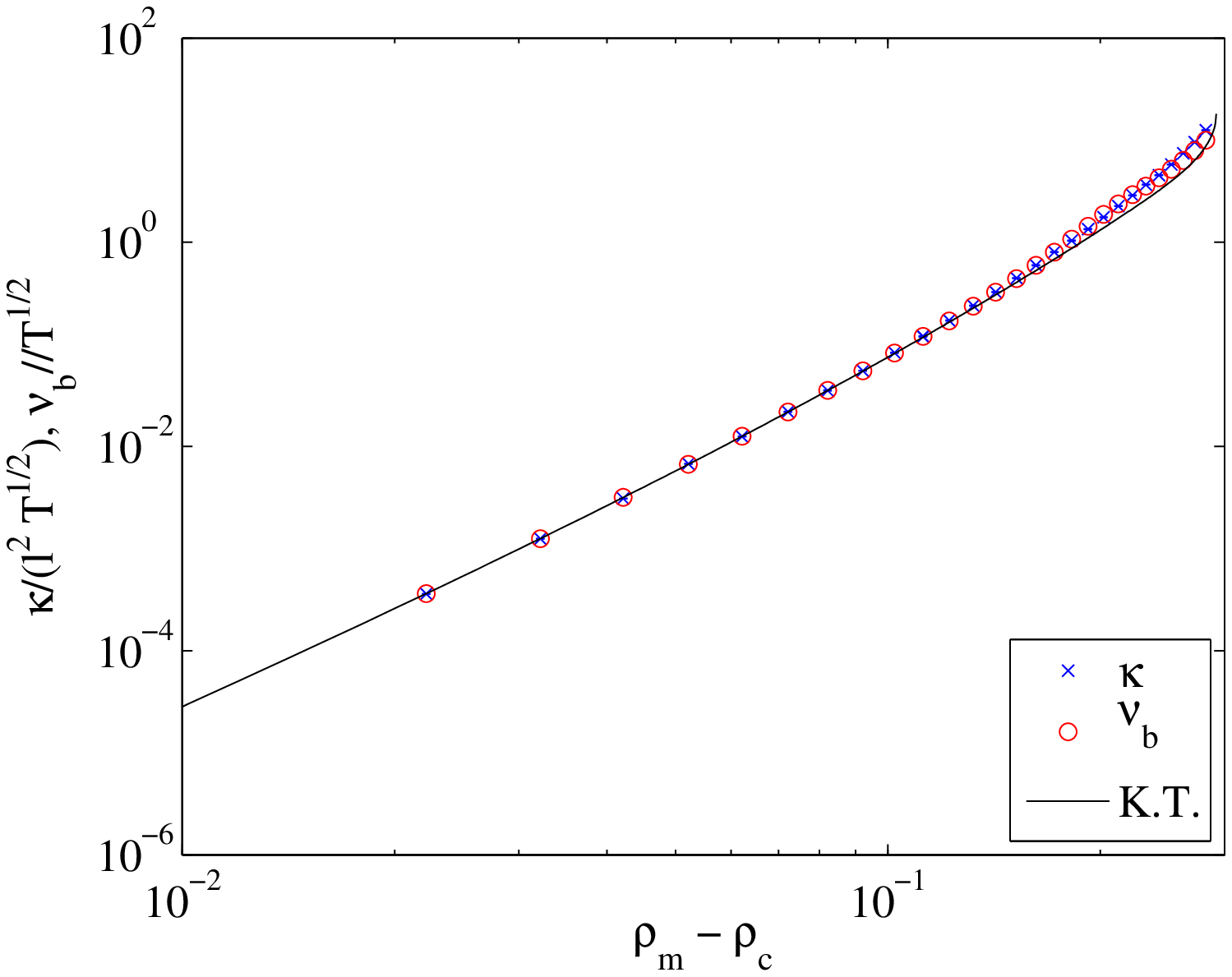}
   \caption{Reduced thermal conductance $\kappa$, computed from the heat
     exchange in a chain with a single particle with thermalized 
     neighbours, 
    and binary collision frequency $\nub$, as
     functions of $\rhom - \rhoc$. (Left) ratio between $\kappa$ and $\nub$;
     (Right) comparison with the results of section \ref{sec.kt}. The only
     relevant parameter is $\rho = 9/25$. For each value of $\rhom$,
     several temperature differences $\delta T$ were taken, all giving
     consistent values of $\kappa$. The solid line shows the result of
     kinetic theory (K.T.).}
   \label{fig.kappa}
\end{figure}

Figure \ref{fig.kappa} shows the results of the computations of the  
heat conductance \footnote{Here and in the sequel, the thermal conductance
  or conductivity are further divided by
$l^2\sqrt{T}$, where $l=1/\sqrt{2}$ is the rhombic cell size, so as to
eliminate its length and temperature dependences, thus defining the reduced
thermal conductivity $\kappa_* = \kappa/(l^2\sqrt{T})$. In these
expressions and from here on, we further set $\kb\equiv1$.}
with this method  and provides a comparison with the
binary collision frequency $\nub$ on the one hand (left panel), as well as
with the results of our kinetic theory predictions on the other hand (right
panel).

The agreement between the data and equations (\ref{nubresult}),
(\ref{heatconductivity}), (\ref{kappanub}) and the computation of the
integrals (\ref{area})-(\ref{A2}), especially as $\rhom\to\rhoc$,
demonstrates the validity of the stochastic description of the billiard
system, equation (\ref{masterequation}).

Next, we increase the size of the chain in order to reach the thermal
conductivity in the limit of an arbitrarily large chain. The results are
that statistical correlations appear between  the kinetic energies along
the chain. As we show below, their influence on the computed value of the
conductivity diminishes as $\rhom\to\rhoc$.

Fix $T_- = 0.5$ and $T_+=1.5$ to be the baths temperatures, and let the
size of the system increase from $N=1$ to $N=20$ as $\rhom$ is
progressively decreased from $\rhom = 11/50$ to $\rhom = 13/100$, with
fixed $\rho = 9/25$. As one can see from figure \ref{fig.kappa}, this range
of values of $\rhom$ crosses over from a regime where the separation of
time scales is not effective to one where it appears to be and where the
stochastic model should therefore be a reasonable approximation to the 
process of energy transport in the billiard. 

For all values of $\rhom$, we measured the temperature  
profile and heat fluxes throughout the system and inferred the value of the
reduced thermal conductivity by linearly extrapolating the ratios between
the average heat flux and local temperature gradient divided by the square
root of the local temperature as functions of $1/N$ to the vertical axis
intercept, corresponding to $N=\infty$. Given the parameter values, every
realisation was carried out 
over a time corresponding to 1,000 interactions between the system and baths
and repeated over $10^4$ realisations. For $N$ up to 20, this time provided
satisfying stationary statistics, with temperature profiles verifying
equation (\ref{noneqTn_nl}) and statistically constant heat fluxes.

\begin{figure}[thp]
   \centering{
     \includegraphics[width = .45\textwidth]{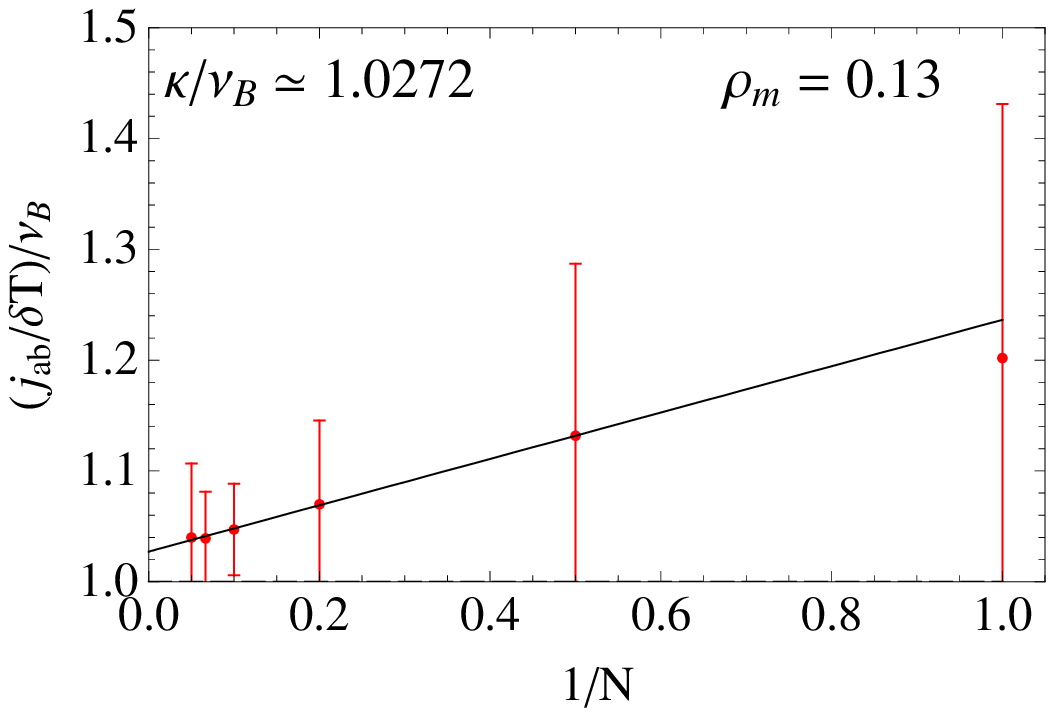}
     \includegraphics[width = .45\textwidth]{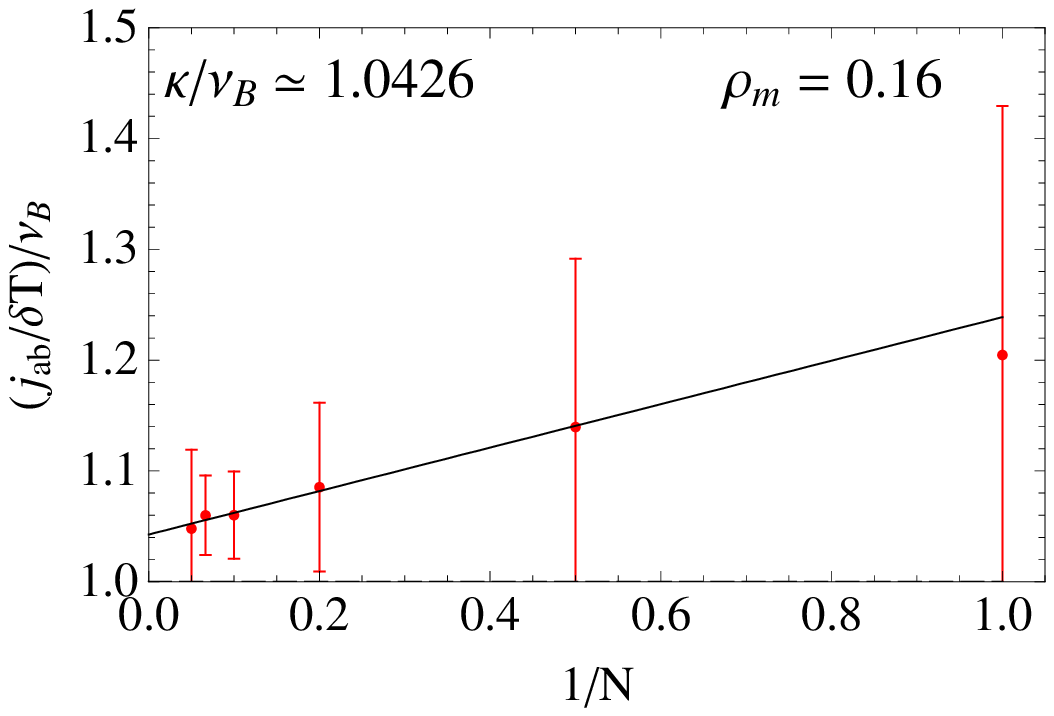}
     \includegraphics[width = .45\textwidth]{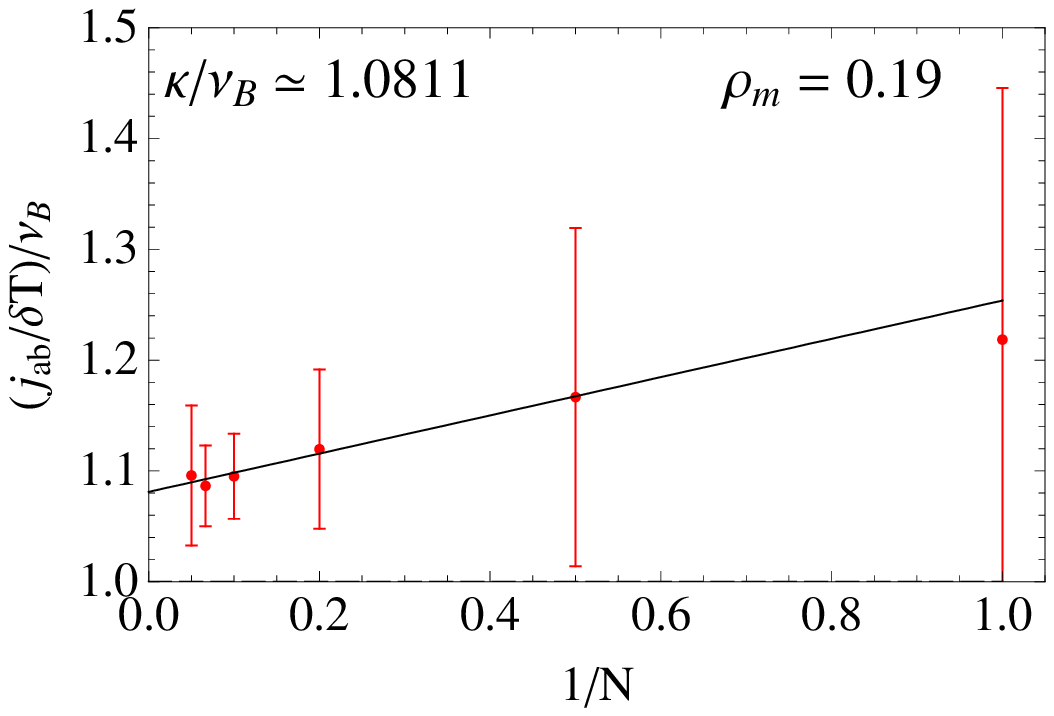}
     \includegraphics[width = .45\textwidth]{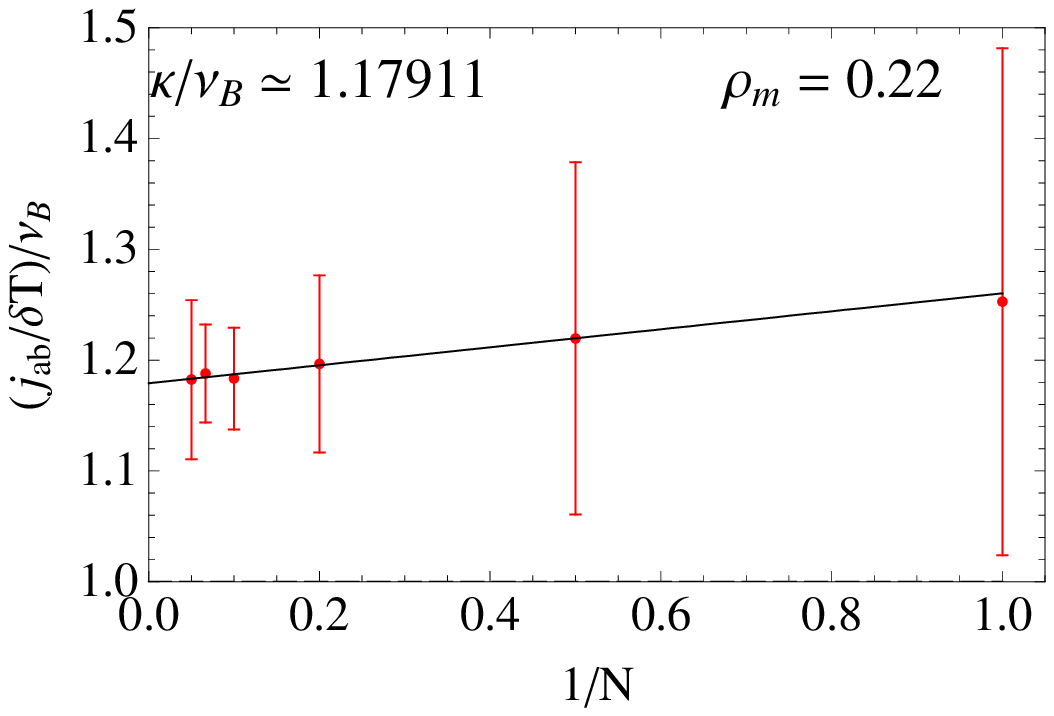}}
   \caption{Ratio between thermal conductivity and binary collision frequency,
     $\kappa/\nub$, extrapolated from the computation of the average ratio
     between heat current and local temperature gradients,  divided by the
     local binary collision frequency,  
     $1/n\sum_i[J_{i,i+1}/(T_{i+1}-T_i)]/\nub(i)$. The system sizes are
     $N=1, 2, 5, 10, 15, 20$. The four pannels correspond to different
     values of $\rhom = 13/100, 4/25, 19/100, 11/50$, with $\rho =
     9/25$ and thus $\rhoc \simeq 0.068$. The red dots are the
     data points with corresponding error bars 
     and the black solid line shows the result of a linear regression
     performed with data associated to systems of lengths $N\ge2$.}
   \label{fig.kappan}
\end{figure}

\begin{figure}[thp]
   \centering{
     \includegraphics[width = .55\textwidth]{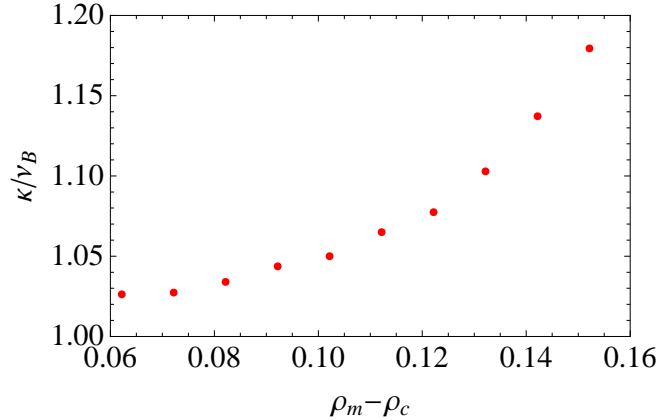}}
   \caption{Ratio between thermal conductivity and binary collision frequency,
     $\kappa/\nub$, computated as in figure \ref{fig.kappan}, here
     collected for a larger set of values of $\rhom$. The horizontal axis
     shows the difference $\rhom-\rhoc$. The error bars are of the same
     order as those in figure \ref{fig.kappa}, with deviations from unity
     of the data points of the same order as those in the latter figure.}
   \label{fig.kappancomb}
\end{figure}

The results of the linear regression used to compute the value of
$\kappa/\nub$ for selective values of $\rhom$ are shown in figures
\ref{fig.kappan} and \ref{fig.kappancomb}.

Our results thus make it plausible that the ratio between the thermal
conductivity and binary collision frequency approaches unity as the
parameter $\rhom$ decreases towards the critical value $\rhoc$. As will be
show in a separate publication \cite{GG08b}, this is indeed a property of
the stochastic model described by equation (\ref{masterequation}) and has
been verified numerically by direct simulation of the master equation
within an accuracy of 4 digits. As of the billiard system, it is
unfortunately difficult to improve the results beyond those presented here
as the CPU times necessary to either increase $N$ or decrease $\rhom$
quickly become prohibitive. Nevertheless, the data displayed in figures
\ref{fig.kappan} and \ref{fig.kappancomb} offer convincing evidence that
the thermal conductivity is well approximated by the binary collision
frequency so long as the separation of time scales between wall and binary
collision events is effective. 

\subsubsection{Helfand moment.}

The computation of the thermal conductivity can also be performed in the
global equilibrium microcanonical ensemble using the method of Helfand
moments \cite{H60,AGW70,GRLB06}.

The Helfand moment has expression $H(t) = \sum_a x_a(t)
\epsilon_a(t)$, where $x_a(t)$ denotes the horizontal position of particle
$a$ at time $t$ and $\epsilon_a(t) = |\bi{v}_a(t)|^2/2$ its kinetic
energy (the masses are
taken to be unity). The computation of the time evolution of this quantity
proceeds by discrete steps, integrating the Helfand moment from one
collision event to the next, whether between a particle and the walls of
its cell, or between two particles. Let $\{\tau_n\}_{n\in\mathbb{Z}}$
denote the times at
successive collision events. In the absence of binary collisions,
the energies are locally conserved and the Helfand moment changes according
to $H(\tau_{n}) = H(\tau_{n-1}) + \sum_a [x_a(\tau_{n}) - x_a(\tau_{n-1})]
\epsilon_a(\tau_{n-1})$. If, on the other hand, a binary collision occurs
between particles $k$ and $l$, the Helfand moment changes by an additional
term $[x_k(\tau_n) - x_l(\tau_n)][\epsilon_k(\tau_{n}+0) -
\epsilon_k(\tau_{n}-0)]$. Computing the time average of the squared Helfand
moment, we obtain an expression of the thermal conductivity according to
\begin{equation}
\kappa = \lim_{L\to\infty} \frac{1}{L(\kb T)^2}
\lim_{n\to\infty} \frac{1}{2\tau_n}
\Big\langle [H(\tau_n) - H(\tau_0)]^2 \Big\rangle
\label{helfand}
\end{equation}
where $L = N/2$ is the horizontal length of the system.

\begin{figure}[thp]
   \centering
   \includegraphics[width = .55\textwidth]{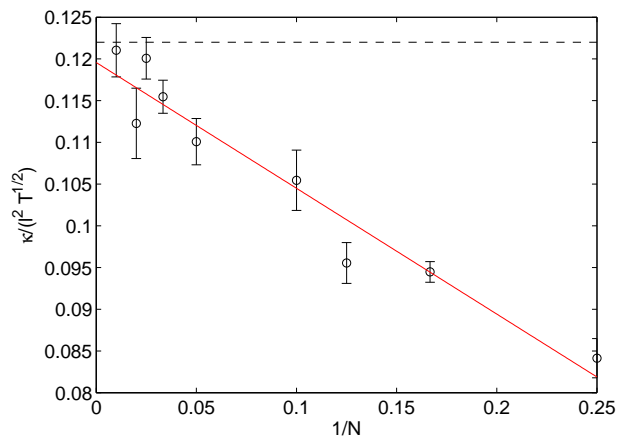}
   \caption{Reduced thermal conductivity, computed from the mean squared
     Helfand moment, versus $1/N$. The parameters are $\rho = 9/25$, $\rhom
     = 9/50$. The system sizes vary from $N=4$ to $N=100$. The dashed line
     shows the binary collision frequency, $\nub \simeq 0.1225$. The solid
     line shows a linear fit of the data, with $y$-intercept $0.12\pm0.01$,
     in agreement with the prediction (\ref{kappanub}).}
   \label{fig.helfand}
\end{figure}

Figure \ref{fig.helfand} shows the results of a computation of the thermal
conductivity through equation (\ref{helfand}) for different system
sizes. Though the actual values of $\kappa$ vary wildly with $N$, it is
clear that a finite asymptotic value is reached for $N\simeq 10^2$.  
In this case, the constant of proportionality in equation (\ref{kappanub})
takes the value $A= 0.98 \pm 0.08$,  close to 1. Similar results were
obtained for other parameter values, and other cell geometries as well.

\section{Lyapunov spectrum \label{sec.ls}}

A key aspect of our model, which justifies the assumption of local
equilibrium, is that it is strongly chaotic. This property
can be illustrated through the computation of the Lyapunov spectrum and
Kolmogorov-Sinai entropy in equilibrium conditions.

As mentioned earlier, in the absence of interaction between the cells,
$\rhom<\rhoc$, The Lyapunov spectrum of a system of $N$ cells has $N$
positive and $N$ negative Lyapunov exponents, which, if divided by the
average speed of the particle to which they are attached, are all equal in
absolute value. This reference value we denote by $\lambda_+$. The $2N$
remaining Lyapunov exponents vanish.

As we increase $\rhom$ and let the particles interact, we expect that, in
the regime $0<\rhom-\rhoc\ll1$, where binary collision events are rare,
the Lyapunov exponents will essentially be determined by $\lambda_+$
multiplied by a factor which is specified by the particle velocities. The
exchange of velocities thus produces an ordering of the
exponents which can be computed as shown below. We note that the other half
of the spectrum, which remains zero in this approximation, will only pick up
positive values as a result of the interactions.

\begin{figure}[thp]
  \centering{
    \includegraphics[width = .8\textwidth]{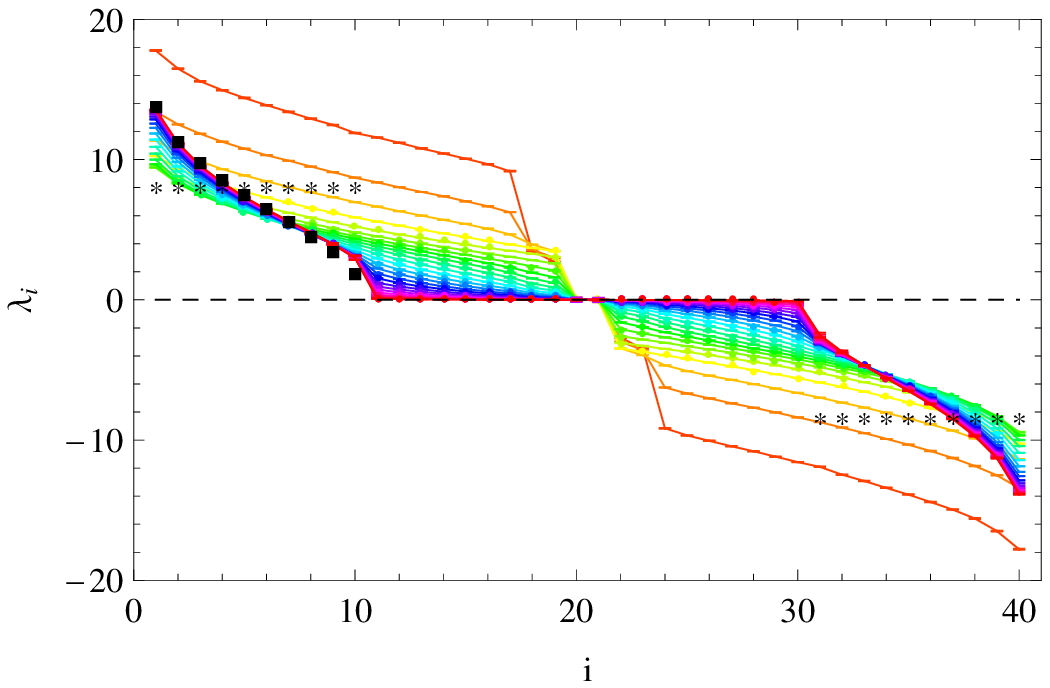}
  }
  \vskip .5cm
  \centering{
    \includegraphics[width = .4\textwidth]{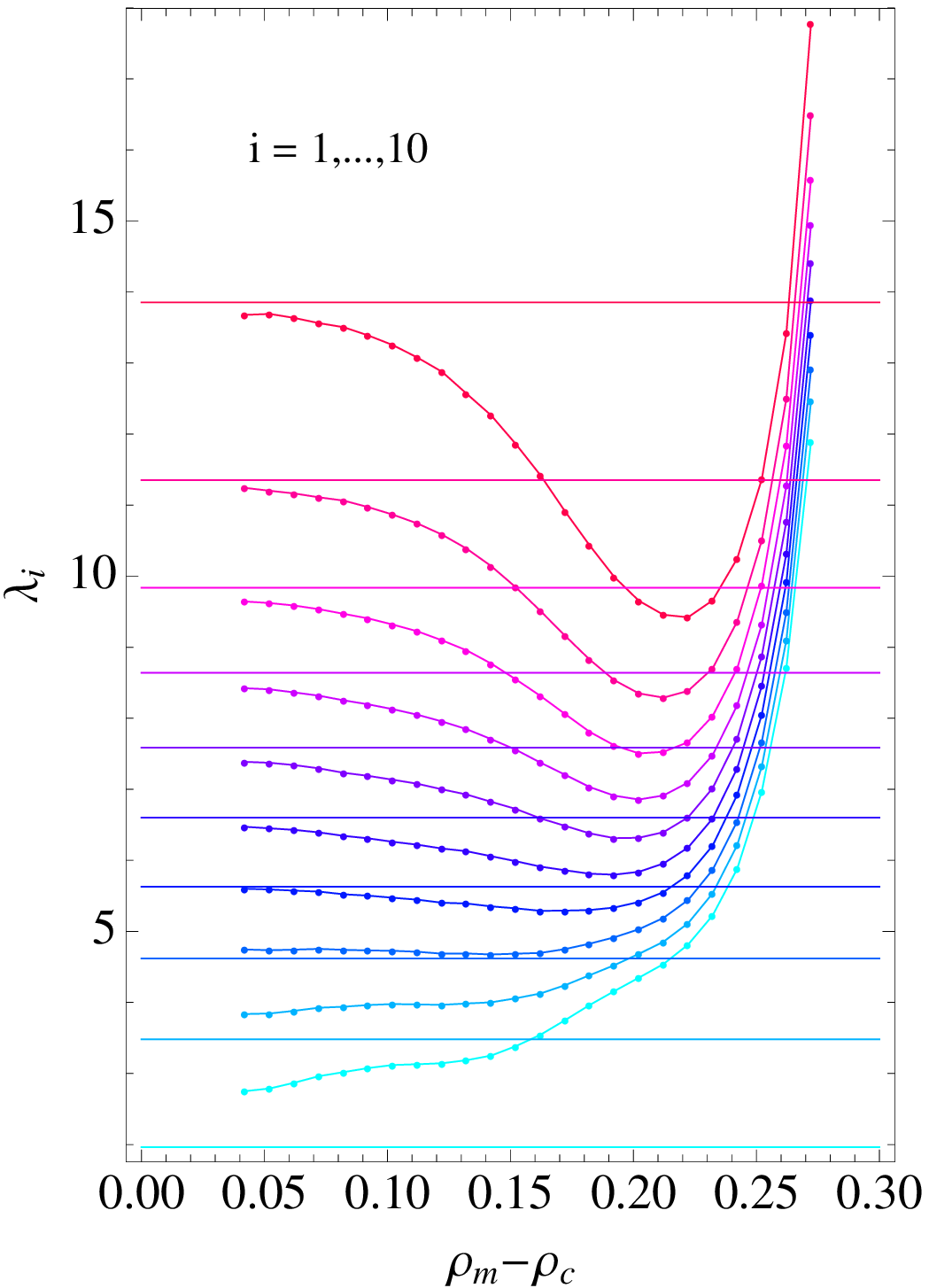}
    \hskip .5cm
    \includegraphics[width = .4\textwidth]{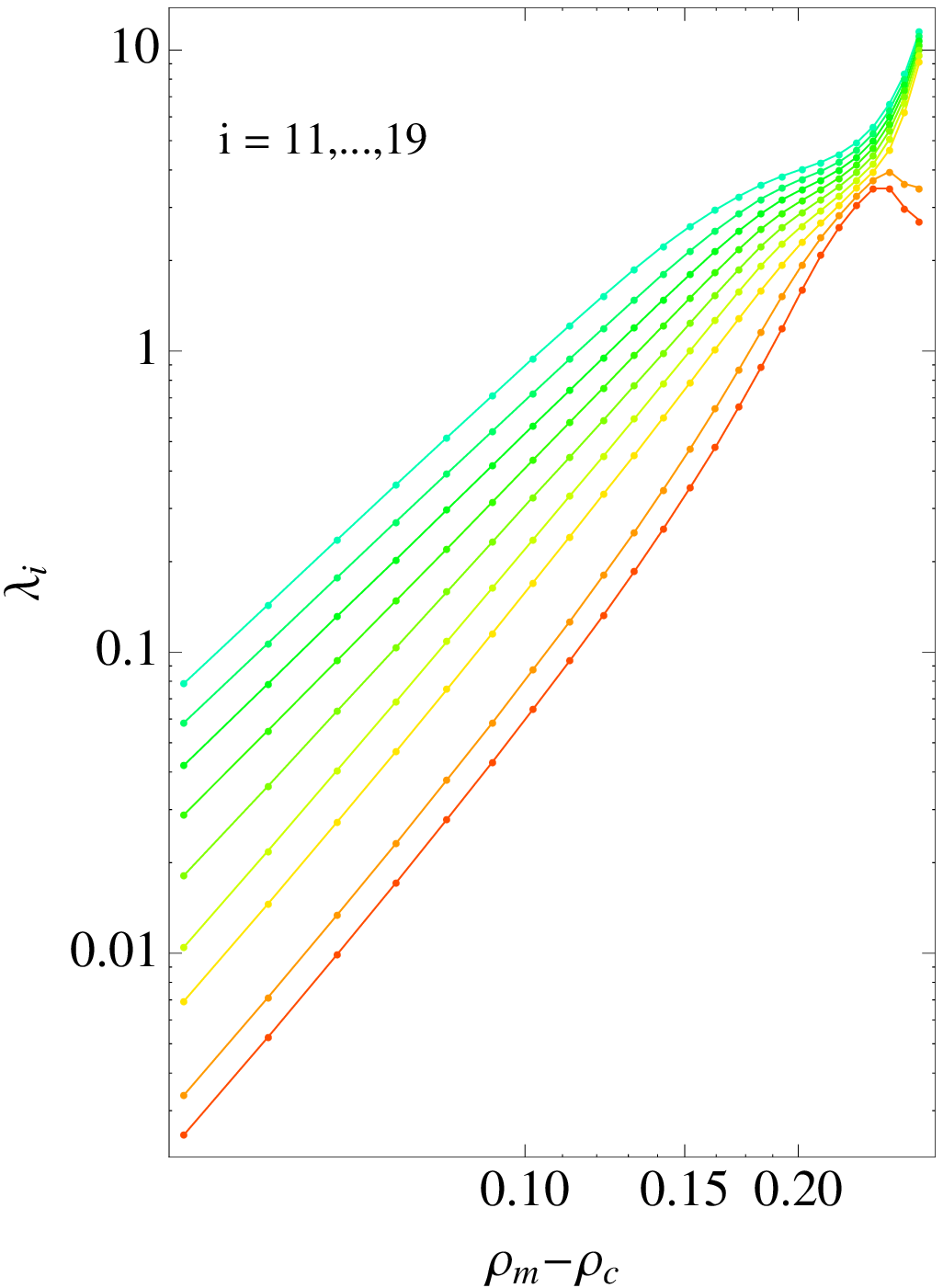}
  }
  \caption{(Top) Lyapunov exponents $\lambda_i$ versus $i$ 
    computed with a rhombic channel of size $N=10$ cells and parameter $\rho = 9/25$.
    The different curves correspond to different values of 
    $\rhom = 0.11$ (bottom curve) to $0.34$ (top curve) by steps of $0.01$.
     The lines of stars are obtained for $\rhom = 0.04 < \rhoc$, 
     yielding the exponents $\lambda_+$ and $\lambda_-$ associated to
     isolated cells, with all the particles at the same speed. The squares
     correspond to the first half of the spectrum as predicted by
     equation (\ref{lambdaimic}). (Bottom) $\lambda_i$ versus
     $\rhom-\rhoc$. The
     first of the two figures displays the first half of the positive part
     of the spectrum of exponents, $\lambda_1,\dots,\lambda_N$ and compares
     them to the asymptotic estimate equation (\ref{lambdaimic}) (straight
     lines). The second plot shows the second half of the positive part of
     the spectrum, $\lambda_{N+1},\dots,\lambda_{2N-1}$, displaying their
     power-law scaling to zero as $\rhom\to\rhoc$.}
   \label{fig.lyap}
\end{figure}

\begin{figure}[thp]
   \centering
   \includegraphics[width = .55\textwidth]{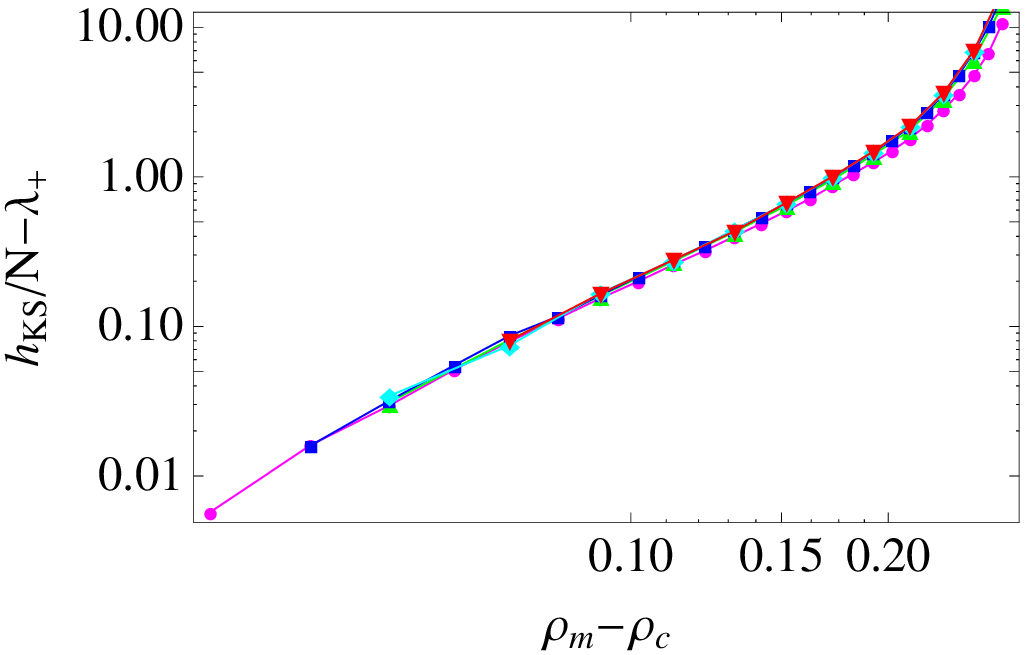}
   \caption{Extensivity of the Kolmogorov-Sinai entropy. The vertical axis
     shows the difference between the Kolmogorov Sinai entropy, here computed from
     the sum of the positive Lyapunov exponents for system sizes $N =3$
     (magenta circles), $5$ (blue up triangles), $10$ (green squares),
     $15$ (cyan diamonds), $20$ (red down triangles),
     divided by the corresponding microcanonical average velocity, and the
     Lyapunov exponent of the isolated billiard cell at unit velocity. The
     curves fall nicely upon each other and converge to zero as
     $\rhom\to\rhoc$.} 
   \label{fig.hks}
\end{figure}

Assume for the sake of the argument that $N$ is large. The probability that
a given particle with velocity $v$ has exponent $\lambda = v \lambda_+$
less than a value $\lambda_i = v_i\lambda_+$ can be approximated by the
probability that the particle velocity be less than $v_i$, which, if we
assume a canonical form of the equilibrium distribution, is
\begin{eqnarray}
   \Pro(\lambda < \lambda_i) &= \Pro(v < v_i),\nonumber\\
   &= \beta\int_0^{m v_i^2/2} \ud \epsilon
   \exp(-\beta \epsilon),\nonumber\\
   &= 1 - \exp(-\beta m v_i^2/2).
\end{eqnarray}
But this probability is simply $(N - i + 1/2)/N$. Therefore the half of the
positive Lyapunov exponent spectrum,
which is associated to the isolated motion of
particles within their cells, becomes, in the presence of rare collision
events,
\begin{equation}
\lambda_i = \lambda_+ \sqrt{\frac{2}{m\beta}}\left[\ln\frac{N}{i-1/2}
\right]^{1/2},
\quad i = 1,\dots,N,
\label{lambdai}
\end{equation}
with ordering $\lambda_1 > \lambda_2>\dots>\lambda_N$. In particular, the
largest exponent $\lambda_1$ grows like $\sqrt{\ln N}$. The
Kolmogorov-Sinai entropy on the other hand is extensive~: $\hks = N
\lambda_+ \sqrt{\pi/(2m\beta)}$.

Refined expressions can be computed by taking the microcanonical
distribution associated to a finite $N$. In particular, the expressions of
the Lyapunov exponents become
\begin{equation}
\lambda_i  = \lambda_+\sqrt{\frac{2 N}{m\beta}}\left[1 -
\left(\frac{i-1/2}{N}\right)^{1/(N-1)}\right]^{1/2}.
\label{lambdaimic}
\end{equation}

We mention in passing that similar arguments are relevant and can be used
to approximate the Lyapunov spectrum (actually half of it) of other models,
such as a mixture of light and heavy particles \cite{GvB02}.

Let us again insist that equations (\ref{lambdai}) and (\ref{lambdaimic})
account for only $N$ of the $2N-1$ positive Lyapunov exponents. $N-1$ are
vanishing within this approximation. For all these Lyapunov exponents, we
expect the corrections to vanish with the binary collision frequency as we
go to the critical geometry.

Figures \ref{fig.lyap} and \ref{fig.hks} show the results of a numerical
computation of the whole spectrum of Lyapunov exponents and corresponding
Kolmogorov-Sinai entropy for the one-dimensional channel of rhombic
cells. The agreement with equation (\ref{lambdaimic}) as $\rhom\to\rhoc$ is
very good, 
at the exception perhaps of the last few among the first $N$ exponents,
whose convergence to the asymptotic value (\ref{lambdaimic}) appears to be
slower. 
Interestingly, the largest exponents have a minimum  which
occurs at about the value of $\rhom$ for which the binary and wall collision
frequencies have ratio unity, see figure \ref{fig.nub}. Indeed, for larger
radii $\rhom$, the spectrum is similar to that of a channel of hard discs
(without obstacles) \cite{FMP04}. Note that the same holds for the ratio
between the thermal conductivity and binary collision frequency, as seen
from figure
\ref{fig.kappa}. Thus one can interpret the occurrence of a minimum of the
largest exponent
as evidence of a crossover from
a near close-packing solid-like phase ($\rhom\lesssim\rho$) to a
gaseous-like phase trapped in a rigid structure ($\rhom\gtrsim\rhoc$).

\section{Conclusions \label{sec.con}}

To summarize, lattice billiards form the simplest class of Hamiltonian
models for which one can observe normal transport of energy, consistent
with Fourier's law. Geometric confinement restricts the transport
properties of this system to heat conductivity alone, thereby avoiding the
complications of coupling mass and heat transports, which are common to
other many particle billiards.

The strong chaotic properties of the isolated billiard cells warrant, in a
parametric regime where interactions among moving particles are seldom, the
property of local equilibrium. This is to say, assuming that wall
collision frequencies are order-of-magnitude larger than binary collision
frequencies, that one is entitled to making a Markovian approximation
according to which phase-space distributions are spread over individual
cells. We are thus allowed to ignore the details of the distribution at the
level of individual cells and coarse-grain the phase-space distributions
to a many-particle energy distribution function, thereby going from the
pseudo-Liouville equation, governing the microscopic statistical evolution,
to the master equation, which accounts for the energy exchanges at a
mesoscopic cell-size scale and local thermalization. The energy exchange
process further drives the  relaxation of the whole system to
global equilibrium.

This separation of scales, from the cell scale dynamics, corresponding to the
microscopic level, to the energy exchange among neighbouring cells at the
mesoscopic level, and to the relaxation of the system to thermal equilibrium
at the macroscopic level, is characterized by three different rates.
The process
of relaxation to local equilibrium has a rate given by the wall collision
frequency, much larger than the rate of binary collisions, which characterizes
the rate of energy exchanges which accompany the relaxation to local thermal
equilibrium, itself much larger than the hydrodynamic relaxation rate, given by
the binary collision rate divided by the square of the macroscopic length of
the system.

On this basis, having reduced the deterministic dynamics of the many
particles motions to a stochastic process of energy exchanges between
neighboring cells, we are able to derive Fourier's law and the macroscopic
heat equation. 

The energy transport master equation can be solved with the result, to be
presented elsewhere \cite{GG08b}, that the binary
collision frequency and heat conductivity are equal. Under the assumption
that the wall and binary collision time scales of the billiard are
well separated, the transposition of this result to the billiard dynamics
is that~: 
\begin{enumerate}
\item The heat conductivity of the mechanical model is proportional to the
  binary collision frequency, {\em i.~e.} the rate of collisions among
  neighbouring particles, 
   \[
   \frac{\kappa}{l^2 \nub} = A,\quad \nub\ll\nuw,
   \]
   with a constant $A$ that is exactly $1$ at the critical geometry,
   $\rhom\to\rhoc$, where the evolution of probability densities is
   rigorously described by 
   the master equation,  and remains close to unity over a large range of
   parameter values for which we conclude the time scale separation is
   effective and the master equation therefore gives a good approximation
   to the energy transport process of the billiard~; 
\item The heat conductivity and the binary collision frequency both vanish
   in the limit of insulating system, $\rhom\to\rhoc$, with $(\rhom -
   \rhoc)^3$,
   \[
   \lim_{\rhom\to\rhoc} \frac{\kappa}{l^2(\rhom - \rhoc)^3}
   =   \lim_{\rhom\to\rhoc} \frac{\nub}{(\rhom - \rhoc)^3}
    = \frac{2\rhom}{|\mathcal{B}_\rho|^2}\sqrt{\frac{\kb T}{\pi m}} c_3,
   \]
   where the coefficient $c_3$ depends on the specific geometry of the
   binary collisions.
\end{enumerate}

Though both results are exact strictly speaking only in the limit
$\rhom\to\rhoc$, the first one, according to our numerical computations, is
robust and holds throughout the range of parameters for which the binary
collision frequency is much less than the wall collision frequency,
$\nub\ll\nuw$. The deviations from $A=1$ which we observed at intermediary
values of $\rhom$, where the separation of time scales is less effective,
are interpreted as actual deviations of the energy transport process of the
billiard from that described by the master equation, where correlations
between the motions of 
neighboring particles must be accounted for. 

Under the conditions of local equilibrium, the Lyapunov spectrum has
a simple structure, half of it being determined according to random
velocity distributions within the microcanonical ensemble, while the
other half remains close to zero. The analytic expression of the Lyapunov
spectrum that we obtained is thus exact at the critical geometry. The
Kolmogorov-Sinai entropy is equal to the sum of the positive Lyapunov
exponents and thus determined by half of them. It is extensive in the
number of particles in the system, whereas the largest Lyapunov exponent
grows like the square root of the logarithm of that number. As we mentioned
earlier, we believe our method is relevant to the computation of the
Lyapunov spectrum of other models of interacting particles
\cite{GvB02}. The computation of the full spectrum, particularly regarding
the effect of binary collisions on the exponents, remains an open problem.

\ack
The authors wish to thank D. Alonso, J. Bricmont, J. R. Dorfman,
M. D. Jara Valenzuela, A. Kupiainen, R. Lefevere, C. Liverani, S. Olla and
C. Mej\'{\i}a-Monasterio for fruitful discussions and comments at different
stages of this work. This research is financially supported by the Belgian
Federal Government under the Interuniversity Attraction Pole project NOSY
P06/02 and the Communaut\'e fran\c{c}aise de Belgique under contract ARC
04/09-312. TG is financially supported by the Fonds de la Recherche
Scientifique F.R.S.-FNRS.

\appendix

\section{Computation of the Kernel \label{app.w}}

We provide in this appendix the explicit form of the transition rate $W$,
given by equation (\ref{energytransition}).

We first substitute the two velocity integrals by two angle integrals,
eliminating the two delta functions which involve only the local energies.
\begin{eqnarray}
\fl\int_{\hat{\bi{e}}_{ab}\cdot\bi{v}_{ab}>0}
\hskip -.35cm \ud\bi{v}_a \ud\bi{v}_b
\hat{\bi{e}}_{ab}\cdot\bi{v}_{ab} \
\delta\left(\epsilon_a - \frac{m v_a^2}{2}\right)
\delta\left(\epsilon_b - \frac{m v_b^2}{2}\right)
\delta\left(\eta -
   \frac{m}{2}[(\hat{\bi{e}}_{ab}\cdot\bi{v}_a)^2  -
(\hat{\bi{e}}_{ab}\cdot\bi{v}_b)^2 ]\right)
\nonumber\\
\lo = \frac{\sqrt{2}}{m^{5/2}}\int_{D_+} \ud\theta_a \ud\theta_b
(\sqrt{\epsilon_a} \cos\theta_a - \sqrt{\epsilon_b} \cos\theta_b)
\delta(\eta - \epsilon_a \cos^2 \theta_a + \epsilon_b \cos^2 \theta_b),
\label{intvel}
\end{eqnarray}
where $\theta_{a/b}$ denote the angles of the velocity vectors
$\bi{v}_{a/b}$ with respect to the direction $\phi$ of the relative
position vector joining particles $a$ and $b$, $\hat{\bi{e}}_{ab} =
(\cos\phi, \sin\phi)$,
and the angle integration is performed over the domain $D_+$ such that
$\sqrt{\epsilon_a} \cos\theta_a >  \sqrt{\epsilon_b} \cos\theta_b$.

With the above expression (\ref{intvel}), the explicit $\phi$ dependence
has disappeared
so that we have effectively decoupled the velocity integration from the
integration over the direction of the relative position between the two
colliding particles. We can further transform this expression in terms of
Jacobian elliptic functions as follows.

Let $x_i = \cos\theta_i$, $i=a,b$, in equation
(\ref{intvel}), which becomes
\begin{equation}
\fl\frac{4\sqrt{2}}{m^{5/2}}
\int_{-1}^{1} \frac{\ud x_a}{\sqrt{1 - x_a^2}}
\int_{-1}^{1} \frac{\ud x_b}{\sqrt{1 - x_b^2}}
\theta(\sqrt{\epsilon_a} x_a - \sqrt{\epsilon_b} x_b)
(\sqrt{\epsilon_a} x_a - \sqrt{\epsilon_b} x_b)
\delta(\eta - \epsilon_a x_a^2 + \epsilon_b x_b^2),
\label{intvel2}
\end{equation}
where $\theta(.)$ is the Heaviside step function. We thus have to perform
the $x_a$ and $x_b$ integrations along the line defined by the
argument of the delta function,
\begin{equation}
\eta = \epsilon_a x_a^2 - \epsilon_b
x_b^2,
\label{delta}
\end{equation}
and that satisfies the condition
\begin{equation}
\sqrt{\epsilon_a} x_a >
\sqrt{\epsilon_b} x_b.
\label{theta}
\end{equation}

To carry out this computation, we have to consider the following
alternatives~:

\begin{enumerate}
\item \framebox{$\epsilon_a<\epsilon_b$, $0<\eta<\epsilon_a$}\\
The solution of equation (\ref{delta}) which is compatible with equation
(\ref{theta}) is
\begin{equation}
x_a = \left(\frac{\eta + \epsilon_b x_b^2}{\epsilon_a}\right)^{1/2}.
\label{case1xa}
\end{equation}
Plugging this solution into equation (\ref{intvel2}) and setting the bounds
of the $x_b$-integral to $\pm\sqrt{(\epsilon_a - \eta)/\epsilon_b}$, the
expression (\ref{intvel2}) reduces to (omitting the prefactors)
\begin{equation}
\int_0^{\sqrt{(\epsilon_a - \eta)/\epsilon_b}} \ud x_b
\frac{1}{\sqrt{\epsilon_a - \eta - \epsilon_b x_b^2}\sqrt{1 - x_b^2}}
= \frac{1}{\sqrt{\epsilon_b}} K \left(\frac{\epsilon_a - \eta}{\epsilon_b}
   \right),
\label{case1elliptic}
\end{equation}
where $K$ denotes the Jabobian elliptic function of the first kind,
\begin{equation}
K(m) =
\int_0^{\pi/2} (1 - m \sin^2\theta)^{-1/2}\ud\theta \quad(m<1).
\label{defelliptic}
\end{equation}

Thus the kernel is, in this case,
\begin{equation}
\fl W(\epsilon_a, \epsilon_b | \epsilon_a - \eta, \epsilon_b + \eta) =
\frac{2\rhom}{\pi^2|\mathcal{L}_{\rho,\rhom}(2)|}
  \sqrt{\frac{2}{m \epsilon_b}}
K \left(\frac{\epsilon_a - \eta}{\epsilon_b}
   \right) \int\ud\phi\ud \bi{R}.
\label{case1W}
\end{equation}

\item \framebox{$\epsilon_a<\epsilon_b$, $\epsilon_a - \epsilon_b < \eta
     <0$}\\
This case is similar to case (i), with equation (\ref{case1xa}) replaced by
\begin{equation}
x_b = - \left(\frac{-\eta + \epsilon_a x_a^2}{\epsilon_b}\right)^{1/2}
\label{case2xb}
\end{equation}
and $-1<x_a<+1$.
The expression of the kernel corresponding to this case is therefore
\begin{equation}
\fl W(\epsilon_a, \epsilon_b | \epsilon_a - \eta, \epsilon_b + \eta) =
\frac{2\rhom}{\pi^2|\mathcal{L}_{\rho,\rhom}(2)|}
  \sqrt{\frac{2}{m (\epsilon_b + \eta)}}
K \left(\frac{\epsilon_a}{\epsilon_b + \eta}
   \right) \int\ud\phi\ud \bi{R}.
\label{case2W}
\end{equation}

\item \framebox{$\epsilon_a<\epsilon_b$, $-\epsilon_b < \eta < 
\epsilon_a - \epsilon_b <0$}\\
This case is similar to case (ii), with 
$-\sqrt{(\epsilon_b+\eta)/\epsilon_a}<x_a<+\sqrt{(\epsilon_b+\eta)/\epsilon_a}$.
In this case, the expression of the kernel is given by
\begin{equation}
\fl W(\epsilon_a, \epsilon_b | \epsilon_a - \eta, \epsilon_b + \eta) =
\frac{2\rhom}{\pi^2|\mathcal{L}_{\rho,\rhom}(2)|}
  \sqrt{\frac{2}{m \epsilon_a}}
K \left(\frac{\epsilon_b + \eta}{\epsilon_a}
   \right) \int\ud\phi\ud \bi{R}.
\label{case3W}
\end{equation}

\end{enumerate}
The cases with $\epsilon_a>\epsilon_b$ are obtained from the cases above
with the roles of $a$ and $b$ interchanged and $\eta\to-\eta$.

\section{Collision area near the critical geometry\label{app.coeff}}
The reason why $c_1$ and $c_2$ in equation (\ref{coefficients}) vanish is
that the angle difference is $\mathcal{O}(\rhom-\rhoc)$ and the area
$A_1(\phi) =  \mathcal{O}[(\rhom-\rhoc)^2]$. These quantities are easily
computed.

Let $\rhom = (1+\varepsilon)\rhoc$, $\varepsilon\ll 1$. We have
\begin{eqnarray}
\phi_\mathrm{T} = \frac{2\rhoc}{l}\varepsilon
- \frac{\rhoc}{l}\varepsilon^2 + \mathcal{O}(\varepsilon^3),\\
\phi_\mathrm{M} =  \frac{2\rhoc}{l}\varepsilon
+ \frac{4\rhoc^3}{l^3}\varepsilon^2 + \mathcal{O}(\varepsilon^3),
\label{angexp}
\end{eqnarray}
which indicates that the bounds of the angle integrals appearing in
equation (\ref{area}) are $\mathcal{O}(\varepsilon)$, with  $\phi_\mathrm{M}$
and $\phi_\mathrm{T}$ differing only to $\mathcal{O}(\varepsilon^2)$.

The leading contribution to the integral $\alpha(\rho,\rhom)$ therefore
stems only from the integration of $A_1(\phi)$, which we can compute
explicitly by expanding $\rhom$ about $\rhoc$ and taking into
consideration that $\phi$ is $\mathcal{O}(\varepsilon)$. The result is
\begin{eqnarray}
\int_0^{\phi_\mathrm{M}}A_1(\phi) \ud \phi &\simeq
\int_0^{\phi_\mathrm{T}}A_1(\phi) \ud \phi,\nonumber\\
&\simeq
\int_0^{\phi_\mathrm{T}} \Big(\frac{16 \rhoc^3}{l} \varepsilon^2
- 4 l \rhoc \phi^2\Big) \ud \phi,\nonumber\\
&= \frac{64 \rhoc^4}{3 l^2}\varepsilon^3,
\label{A1lead}
\end{eqnarray}
which yields the leading coefficient $c_3$ in equation (\ref{coefficients}).

We can compute the coefficients of the next few powers in the expansion
(\ref{alphaexp}) in a similar fashion. First we notice that $A_2(\phi)$ is
$\mathcal{O}(\varepsilon^3)$ so that its integral between $\phi_\mathrm{T}$
and $\phi_\mathrm{M}$ is $\mathcal{O}(\varepsilon^5)$. Therefore only
  the integral of $A_1(\phi)$ contributes to $c_4$ in equation
  (\ref{coefficients}).

The computation of the next terms in the
expansion is more involved since it requires the integration of
$A_2(\phi)$, whose expression is~:
\begin{eqnarray}
A_2(\phi) =& 8\rho^2\arcsin\left[\frac{\rhom l \sin\phi - (\rhom^2 -
   \rhoc^2)}{\rho^2}\right]^{1/2}\\
&- 4 [\rhom l \sin\phi - (\rhom^2 -  \rhoc^2)]^{1/2}(
4\rhom^2 + l^2 - 4 \rhom l \sin\phi)^{1/2}\nonumber.
\end{eqnarray}
Expanding this expression for $\rhom$ $\varepsilon$-close to $\rhoc$ and
$\phi$ $\varepsilon^2$-close to $\phi_\mathrm{T}$, we get, to leading
order,
\begin{equation}
\int_{\phi_\mathrm{T}}^{\phi_\mathrm{M}}A_2(\phi)\ud \phi
= \frac{1024 \rho^4\rhoc^4}{15 l^6 }\varepsilon^5.
\end{equation}
Combining this expression with the 5$th$ order contribution to the
integral of $A_1(\phi)$, we obtain $c_5$, equation (\ref{coefficients}).
We point out that this coefficient is actually much larger than $c_3$ and
$c_4$, a reason being that negative powers of $\rhoc$ appear in its
expression. The same holds of the next few coefficients.

\end{document}